\documentclass[twocolumn,aps,prx,amssymb,raggedbottom,nobalancelastpage,superscriptaddress]{revtex4}

\usepackage{amsmath}
\usepackage{amssymb}
\usepackage{amsfonts}
\usepackage{dsfont}
\usepackage{graphicx}
\usepackage{bm}
\usepackage{color}
\usepackage{appendix}
\usepackage{epsfig}
\usepackage{bbm}

\newcommand{\RN}[1]{%
  \textup{\uppercase\expandafter{\romannumeral#1}}%
}

\newcommand{\rn}[1]{%
  \textup{\textbf{\lowercase\expandafter{\romannumeral#1}}}%
}

\newcommand\be{\begin{equation}}
\newcommand\ee{\end{equation}}
\newcommand\bea{\begin{eqnarray}}
\newcommand\eea{\end{eqnarray}}

\newcommand\bbea{\begin{aligned}}
\newcommand\eeea{\end{aligned}}

\newcommand\ket[1]{\left|#1\right\rangle}

\DeclareMathOperator{\Tr}{Tr}
\providecommand{\e}[1]{\ensuremath{\times 10^{#1}}}

\begin{document}
\title{Wavefunctions of Symmetry Protected Topological Phases from Conformal Field Theories}
\author{Thomas Scaffidi}
\affiliation{Rudolf Peierls Centre for Theoretical Physics, Oxford OX1 3NP, United Kingdom}
\author{Zohar Ringel}
\affiliation{Rudolf Peierls Centre for Theoretical Physics, Oxford OX1 3NP, United Kingdom}
%

\begin{abstract}
We propose a method for analyzing two-dimensional symmetry protected topological (SPT) wavefunctions using a correspondence with conformal field theories (CFTs) and integrable lattice models. 
This method generalizes the CFT approach for the fractional quantum Hall effect wherein the wavefunction amplitude is written as a many-operator correlator in the CFT.  
Adopting a bottom-up approach, we start from various known microscopic wavefunctions of SPTs with discrete symmetries and show how the CFT description emerges at large scale, thereby revealing a deep connection between group cocyles and critical, sometimes integrable, models.
We show that the CFT describing the bulk wavefunction is often also the one describing the entanglement spectrum, but not always. Using a plasma analogy, we also prove the existence of hidden quasi-long-range order for a large class of SPTs.
Finally, we show how response to symmetry fluxes is easily described in terms of the CFT.
%
%
%
%
%
\end{abstract}


\maketitle

\section{Introduction}
In the past three decades, topological phases of matter have attracted a large amount of interest due to their tendency to exhibit highly robust quantum phenomena which have various applications in quantum engineering and metrology. One of the current frontiers in the field aims at understanding the variety of novel topological phases which arise when some extra symmetries are not allowed to be broken. For example, it was shown~\cite{Chen2011b,Chen2011,Levin2012} that, given an Ising ($Z_2$) symmetry, there are two topologically distinct Ising paramagnets in two dimensions. This can be thought of as the spin analog of topological insulators~\cite{Hasan2010} and accordingly the topological Ising paramagnet must have gapless magnetic excitations on its boundary. While numerous types of topological insulators have been realized experimentally~\cite{Hasan2010,Rasche2013,story2012}, such ``topological paramagnets" have thus far only been realized in the 1D context~\cite{Buyers1986}, although several ideas have been put forward concerning 2D~\cite{Wen2014,Maciek2015} and 3D~\cite{PhysRevB.91.195131}.

Since this discovery, a variety of topological paramagnets (more commonly known as short-range entangled bosonic SPTs) with different symmetries and different dimensions have been explored using various advanced tools ~\cite{Chen2011,YuanMing2012,Kapustin2014,Schuch2011,Xu2013,PhysRevB.91.134404}. Long-range entangled versions of these phases (coined symmetry enriched topological (SET) phases) have been studied as well \cite{YuanMing2013, WenEnriched2013}.  
Still, open questions remain concerning the scope of these approaches \cite{Kapustin2014}, their extensions to fermionic systems \cite{Gu2015}, and their relevance to experimentally feasible models. Also, microscopic lattice models realizing fractional topological paramagnets~\footnote{These phases would correspond to SETs for which the gauge field is emergent.} (for instance gapped spin systems with an Ising ($Z_2$) symmetry supporting bulk excitations with a $Z_4$ symmetry) are very scarce \footnote{An example of this would be a generalization of the Kalmeyer-Laughlin state\cite{PhysRevLett.59.2095} to lower filling fractions.}. 
%

A powerful and conceptually simple tool in analyzing the fractional quantum Hall effect (FQHE), a prominent 2D topological phase, is the FQHE-CFT correspondence ~\cite{Fubini1991,MooreRead1991,Blok1992615,NayakRevMod2008}. 
While CFTs are usually used to describe critical states, here the ground state wavefunction of a 2D topological phase is expressed as a many-operator correlation function in the holomorphic (chiral) part of a CFT:
\begin{equation}
\ket{\psi_{\text{FQHE}}} =  \sum_{ z_i} \left\langle \prod_i \mathcal{O}(z_i) \right\rangle_{\text{CFT}} \bigotimes_i \ket{z_i},
\label{EqFQH}
\end{equation}
where $z_i$ are the particle positions given on the complex plane and the expectation value is computed in a given CFT in which $\mathcal{O}$ is a given operator.
Choosing different CFTs and different operators $\mathcal{O}$ yields a variety of Abelian, non-Abelian, bosonic, and fermionic fractional quantum Hall (FQH) phases. 
The quasiparticle statistics, ground state degeneracies, and edge spectrum are all readily deduced from known properties of CFTs~\cite{NayakRevMod2008}. Furthermore microscopic Hamiltonians which stabilize these phases can be written down.

There are many recent works showing similarities between SPTs and the FQHE~\cite{YuanMing2012,Zaletel2012,Ryu2014,Ringel2015}. Most relevant to our discussion is Ref. \onlinecite{YuanMing2012}, where SPT phases are analyzed through the prism of bi-layer or multi-layer quantum Hall effects. Using the K-matrix formulation, several classes (though not all~\footnote{Their construction does not generalize straightforwardly to non-Abelian symmetries or to an important subclass of Abelian SPTs which result in a non-Abelian theory when gauged (type $\rn{3}$ cocycles)}) of SPTs were shown to correspond to compact boson CFTs with an unconventional (chiral) action of the physical symmetry on the left and right goers (see also Ref. \onlinecite{Ryu2015}). While this suggests that some ground state wavefunctions of SPT phases can be presented as correlators in these compact boson CFTs, this continuum observation appears physically relevant only for bi-layer quantum Hall setups and not for spins on a lattice. The main microscopic approach to studying SPTs on a lattice is the group cohomology approach\cite{Chen2011b,Chen2011} and, at least for bosonic SPTs, it is also more comprehensive in its scope. Physically, however, it is somewhat opaque and it is in particular unclear what CFTs one can associate with these wavefunctions and whether this is beneficial in some way.

Our work therefore provides a distinct and unifying approach to studying SPT phases which generalizes the above CFT-FQHE correspondence and at the same time interpolates between the continuum field theory approaches of Refs. \cite{YuanMing2012,PhysRevB.91.134404} and the microscopic group cohomology approach. 
It is shown that microscopic 2D bosonic SPT wavefunctions, when written in the symmetry-charge basis (i.e. the basis on which the symmetry acts diagonally), appear as a many-operator correlation function in a non-chiral CFT (i.e. where both holomorphic and anti-holomorphic parts are taken). Based on loop models and notions of discrete flux attachment \cite{Ringel2015}, a microscopic toolbox is developed which allows us to identify CFTs associated with a large variety of SPTs including SPTs for which the continuum approaches do not apply in any straightforward way. 
Furthermore, the expected symmetry-flux responses~\cite{Zaletel2012,Ryu2014} are obtained.
For a $Z_{N>2}$ symmetry, this CFT approach is used to establish the presence of hidden order~\cite{Quella2013,Else2013,Ringel2015}, a property which to the best of our knowledge has not been derived before. Somewhat surprisingly, unlike in the CFT-FQHE context where it is believed to be impossible to obtain an exact tensor product state (TPS) description, here there appears to be no tension between having a TPS and writing the wave function as correlators in a CFT. Lastly, we show that for all group cohomology wave functions, the entanglement spectrum is given by the spectrum of the CFT.   



\section{Microscopic SPT-CFT correspondence}
We consider the ground state $\ket{\psi}$ of a 2D bosonic SPT based on a discrete symmetry group $G$. 
The Hilbert space is given by degrees of freedom $\phi_r \in G$ lying on the sites of a triangular lattice at position $r$. 
For simplicity, let us focus on $G=Z_N$, where we label group elements as the numbers $\{0,\dots, N-1\} \simeq G$. 
%
In the basis $\ket{\phi_r}$, the action of  $g \in G$ is given by $g |\phi_r\rangle = |g+\phi_r\rangle$. 
Since it rotates under the action of $G$, we refer to this basis as the symmetry-phase basis.
The SPT wavefunction is then given by $|\psi \rangle = \mathcal{N}^{-1/2}\sum_{\{ \phi_r \}} A_{\{ \phi_r \}} | \{ \phi_r\} \rangle$ where all the $\mathcal{N}$ configurations $\{\phi_r\}$ are summed over.
%
For all known wavefunctions, $A_{ \{\phi_r\}}$ is given by a product of local phase factors \cite{Levin2012,LevinWen2005}.


To establish similarities with FQH wavefunctions, which are written as a function of the charge positions (the charge being electrons in that case), it is natural to use the symmetry-charge basis, $\alpha_r \in \{0,\dots, N-1\}$, on which $g$ acts diagonally: $g | \alpha_r \rangle = e^{2 \pi i \alpha_r g/N} | \alpha_r \rangle$~\cite{Ringel2015}.
For non-Abelian $G$, this would be the representation basis.
It will be advantageous to enumerate the $\{\alpha_r\}$ basis by a set of unordered positions $\{\pm_i, r_i\}$ of ``charge $\pm 1$ particles'' such that $\alpha_r = \sum_{i} \pm_i \delta_{r,r_i} \mod N$. 
Redundancy in this description is removed by always taking the smallest number of particles.
The SPT wavefunction reads
\begin{equation}
\begin{aligned}
\label{Eq:SPTCFT}
\ket{\psi} &= \frac{Z}{\mathcal{N}} \sum_{\{ \pm_i,r_i\}} A_{ \{ \pm_i,r_i\}} \ket{\{ \pm_i,r_i\}} \\ 
A_{ \{ \pm_i,r_i\}} &= Z^{-1} \sum_{\{ \phi_r\}}A_{ \{ \phi_r\}} \prod_i e^{\mp_i 2 \pi i \phi_{r_i}/N} \\ 
Z &\equiv \sum_{\{ \phi_r\}}A_{ \{ \phi_r\}}
\end{aligned}
\end{equation}
The amplitudes of $|\psi\rangle$ in the symmetry-charge basis appear as correlators in a statistical mechanics model whose configuration space is $\{ \phi_r \}$, whose ``Boltzmann" weights are $A_{\{ \phi_r \}}$, and where an operator $O_{\pm}(r) \equiv e^{\mp 2 \pi i \phi_{r}/N}$ is inserted at every charge $\pm 1$ particle position. 

A major motivation for our approach is the observation that two-point correlation functions within this partition function, also known as strange correlators between the SPT state and an ideal paramagnet, appear to be critical~\cite{Cenke2014}.
In the following, we will show explicitly for many examples that this $Z$ theory is indeed critical, and that it also gives the entanglement spectrum theory. Even so, for Eq.~\ref{Eq:SPTCFT} to be a useful generalization of Eq.~\ref{EqFQH}, three main issues must be addressed: (a) the identification of the CFT governing this critical behavior, (b) the association between $O_{\pm}$ and CFT operators, and (c) the strong ultraviolet corrections to CFT predictions, coming from the fact that the $O_{\pm}$ operator insertions are dense on the lattice scale.


To address the last issue, it will be advantageous to consider the dilute symmetry charge regime where the number of particles per lattice site $\rho$ is made small by adding a particle fugacity factor to the wavefunction. 
In this regime, the correlation function $A_{ \{ \pm_i,r_i\}}$ involves mainly long-range features which can therefore be studied using CFT.
%
Provided that the wavefunction near $\rho=0$ is adiabatically connected to that with $\rho \approx 1$, the qualitative features thus obtained should persist down to the lattice scale ($\rho \approx 1$). 
To establish this adiabatic connection, in Appendix \ref{Sec:NumericalDilute} we use the fact that $\ket{\psi}$ can be written as a tensor product state (TPS) to show that a continuous family of local fugacity-dependent Hamiltonians exists for which these fugacity-dependent wavefunctions are the exact ground states. 
Moreover, these ground states are found to be strictly short-range correlated, strongly implying that this family of Hamiltonians is gapped as required. Note that this adiabatic connection is akin to that between FQH phases in the continuum ($\rho \rightarrow 0$) and their lattice counterparts, coined Fractional Chern Insulators, for which $\rho$ is of order one. Here again, there is ample numerical~\cite{Sorensen2005,Hafezi2007,Scaffidi2012,Liu2013} and analytical~\cite{Kapit2010,Qi2011,Wu2013,Harper2014,Scaffidi2014} evidence.

\section{Entanglement spectrum}
\begin{figure}[t!]
\includegraphics[width=75mm, clip=true]{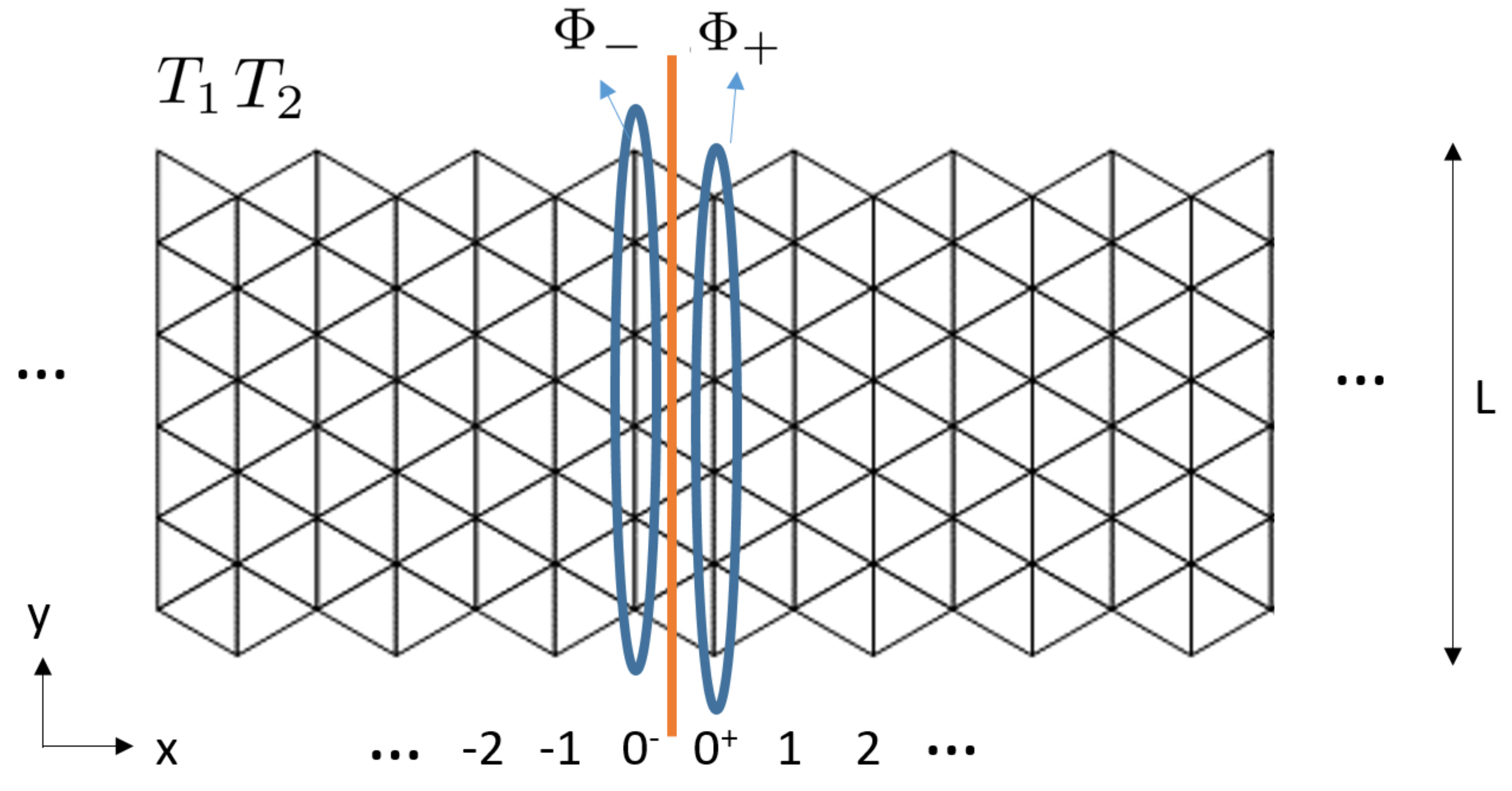}
\caption{Bipartition of the infinite cylinder (infinite in the $x$ direction, periodic boundary conditions along $y$ with width $L$). $\Phi_{\pm}$ is an index running over all possible configurations of $\phi(0^{\pm},y)$.}
\label{Fig:Entanglement}
\end{figure}

We now explicitly show the relation between the theory describing the SPT bulk wavefunction ($Z$) and the theory describing the entanglement spectrum. 
While it is difficult to extract this spectrum in general, a significant simplification arises from an important feature of all the ``fixed-point wave functions'' \cite{LevinWen2005} we consider in this work: $A_{ \{ \phi_r\}}$ always appears as a product over triangles of local phase factors. 
The partition function for a triangular lattice on a torus of length $N$ and width $L$ (see figure \ref{Fig:Entanglement}) can therefore be written as 
\begin{align}
Z &= \Tr[(T_1 T_2)^{N/2}],
\end{align}
where $T_1$ and $T_2$ are $|G|^{L} \times |G|^{L}$ transfer matrices for which the matrix elements $T_{\Phi,\Phi'}$ are given by the product of phase factors for a given column of triangles and where the group elements on the left and right side of this column are given by $\Phi$ and $\Phi'$, respectively. 
If $Z$ describes a conformal field theory, the low lying spectrum of $\log(T_1 T_2)$ is, up to a simple normalization, equal to the spectrum of scaling dimensions in the CFT. 

Let us consider a system defined on an infinite cylinder and consider cutting the cylinder in half.
We define $ r^- \in \{(x,y) | x\leq0^-\}$, $r^+ \in \{(x,y) | x\geq0^+\}$, $\phi_{\pm} \in \{ r^{\pm} \rightarrow G\}$.
We compute the reduced density matrix of the left half of the cylinder. 
It is given by
\begin{equation}
\rho\left[\phi_-;\phi'_-\right] = \sum_{\phi_+} \psi\left[ \phi_-, \phi_+  \right] \psi^*\left[ \phi'_-, \phi_+  \right] 
\end{equation}

We now decompose the product of group cocycles into three factors
\begin{equation}
\psi\left[ \phi_-, \phi_+  \right] = A_-[\phi_-] T_{\Phi_{-},\Phi_{+}} A_+[\phi_+] 
\end{equation}
where $\Phi_{\pm}(y) \equiv \phi_{\pm}(0^{\pm},y) $ and where $T$ is either $T_1$ or $T_2$ depending on where the cut lies (we will take $T=T_1$ in the following without loss of generality and in agreement with Figure \ref{Fig:Entanglement}). 
$A_{\pm}[\phi_{\pm}]$ is the product of cocycles over all triangles strictly inside $\{r^{\pm}\}$. %
%

Using the fact that $|A_+[\phi_+]|^2=1$, the cocycle factors disappear for all triangles strictly inside $\{r^+\}$, and one can trivially compute the sum for all sites with $x>0^+$, leading to
\be
\bbea
\rho\left[\phi_-;\phi'_-\right] &= A_-[\phi_-]  A_-^*[\phi'_-]  \sum_{\Phi_+} T_{\Phi_-, \Phi_+} T^*_{\Phi'_-, \Phi_+} \\
&= A_-[\phi_-]  A_-^*[\phi'_-] \  \mathcal{T}_{\Phi_-,\Phi'_-} \\
&= \sum_{\lambda} \lambda \ \psi_{\lambda}(\phi_-) \psi^*_{\lambda}(\phi'_-)
\eeea
\ee
where $\mathcal{T}\equiv TT^{\dagger}$, $\mathcal{T}_{\Phi,\Phi'} \ u_{\Phi'}^{\lambda} = \lambda \ u_{ \Phi}^{\lambda}$ and $\psi_{\lambda}(\phi_-) \equiv A_-[\phi_-] u_{ \Phi_-}^{\lambda}$.
Using the fact that $|A_-[\phi_-]|^2=1$, it is easy to see that the $\psi_{\lambda}$ form an orthonormal set.
We can therefore conclude that the entanglement spectrum $\lambda$ is given by the spectrum of $\mathcal{T}$.

We now have two theories defined by two different transfer matrices: the theory for the bulk wavefunction is given by $T_1 T_2$ and the theory for the entanglement spectrum is given by $T_1 T_1^{\dagger}$.
Now, within the group cohomology construction of SPT wavefunctions, the local cocycle factor is taken with a complex conjugate for, say, all left-pointing triangles.
It is easy to see that this complex conjugation ensures that $T_2 = T_1^{\dagger}$, and that the two theories are therefore the same.

While the bulk and edge theory are the same for group cohomology wavefunctions, they do not have to be the same in general. 
For example, the Levin-Gu wavefunction does not include this complex conjugation, and in that case $T_2 = T_1^t \neq T_1^{\dagger}$.
In that case, the bulk and edge theories are therefore different: as discussed below, while the edge theory ($T_1 T_1^{\dagger}$) describes a free boson with $c=1$, the bulk theory ($T_1 T_1^t$) is a highly non-trivial $c=-7$ non-unitary, logarithmic CFT.
A similar situation was shown to arise for the Haldane-Rezayi Fractional Quantum Hall state \cite{Gurarie:1997dw}.

It is interesting to note that the transfer matrix of the edge theory is always Hermitian, as it should be if it is to describe the edge dynamics, while the transfer matrix for the bulk theory can be non-Hermitian (and the corresponding CFT non-unitary).
Note that in this section, ``edge theory'' refers to the theory giving the entanglement spectrum, as this is a well-defined theory for a given wavefunction, unlike the theory describing a physical edge, which depends on the arbitrary choice of admissible terms added to the edge Hamiltonian.




\section{Identification of CFTs}

After having established the usefulness of $Z$ for stuyding both the bulk wavefunction and the entanglement spectrum, we now establish its criticality and identify the emerging CFT for a large set of examples.
Before delving into a more specific analysis, several guiding principles should be identified. 

The first one concerns symmetries. The microscopic wave functions coming from group cohomology, and consequently also the partition function $Z$, all obey the symmetry group of the SPT $G$ when placed on closed boundary conditions. 
However there is often a large freedom of choice in writing down such wave functions which can lead to extra on-site symmetries. 
As a starting point, and in order to facilitate the calculations, it is natural to make the most symmetric choice available.
In practice we find that this amounts to choosing a particular branching structure with an hexagonal unit cell. 

The second one concerns the relation to a classical Statistical Mechanics (Stat. Mech.) model. 
The partitions functions $Z$ obtained here all have complex Boltzmann weights. 
However, in all examples given below (except the Levin-Gu wavefunction and the case of $G=D_3$ with $p \neq 3$), we find that judiciously summing-out a particular sublattice of the triangular lattice leads us to a true statistical mechanical models with real, positive Boltzmann weights. 
These classical models appear either as loop models or spin-ice like models having zero discrete divergence constraints. 
Besides being amenable to Monte Carlo simulations, in many cases these models can be solved by promoting the discrete degrees of freedom to continuous ones and establishing a suitable map to a compact boson.


Remarkably we find three classical models obtained from group-cohomology wave functions which are in fact integrable models for which the CFT can be obtained in a rigorous manner. In several other cases where we are not aware of mappings to integrable models, our numerical results show some hints of integrability: CFT-implied degeneracies are exact on the lattice and finite size corrections appear particularly small. This raises the intriguing possibility that group cohomology cocycles are linked with integrable models.

\subsection{Abelian Symmetries}
\label{Sec:Type1}
\subsubsection{Type $\rn{1}$ cocycles}

\begin{figure}[t!]
\includegraphics[width=50mm, clip=true]{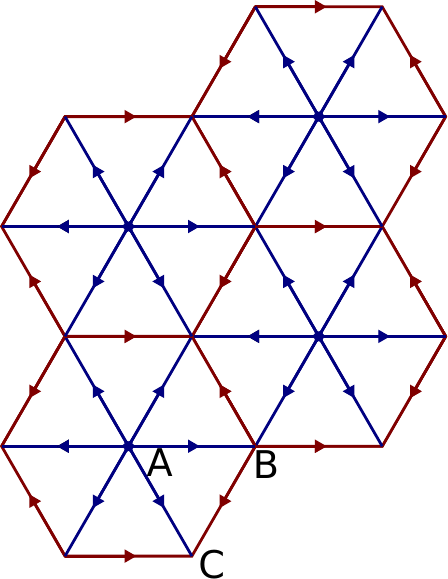}
\caption{Advantageous branching structure for which points on the $A$ sublattice (shown in blue) always appear first in the ordering of vertices in their surrounding triangles. The $B$ and $C$ sublattices are shown in red. In each triangle, the first vertex has two outgoing edges, the second vertex has one outgoing and one incoming edge, and the third vertex has two incoming edges. This branching structure is referred to as the $ABC$ branching structure in the following.}
\label{Fig:BranchingStructure}
\end{figure}

First is the group cohomology wavefunctions for an SPT with a $Z_N$ symmetry~\cite{Chen2011}. 
To write such wavefunctions, one must choose a branching structure, or equivalently, a consistent ordering of the vertices, $(r_1,r_2,r_3)$, of each triangle on the lattice. 
Crucially for later analysis, we choose this ordering such that, for each triangle, the first, second and third vertex belongs to the A, B and C sublattice, respectively (see Fig.~\ref{Fig:BranchingStructure}). In the following, we will refer to this choice as the $ABC$ branching structure.
In 2D, there are $N$ different paramagnetic phases with $Z_N$ symmetry.
If we index them by $p=0,\dots,N-1$, a wavefunction belonging to the $p$-th SPT phase is given by
\begin{equation}
\begin{aligned}
A_{\{ \phi_r \}} &= \prod_{(r_1,r_2,r_3) \in {\cal T}} \nu_p^{s_{(r_1,r_2,r_3)}} (0,\phi_{r_1},\phi_{r_2},\phi_{r_3}) \\ 
\nu_p(0,\phi_{1},\phi_{2},\phi_{3}) &= e^{\frac{2\pi i p}{N} \phi_{1} f(\phi_{1},\phi_{2},\phi_{3})} \\
f(\phi_{1},\phi_{2},\phi_{3}) &= \frac{\langle \phi_{2}-\phi_{1}\rangle+\langle \phi_{3}-\phi_{2}\rangle-\langle \phi_{3}-\phi_{1}\rangle}{N} 
\label{Cocycle}
\end{aligned}
\end{equation}
where ${\cal T}$ is the set of triangles with ordered vertices indexed by $(r_1,r_2,r_3)$, $s_{(r_1,r_2,r_3)}$ is $1$ ($-1$) if the vertices are clockwise (anticlockwise) oriented, $\nu_p$ is known as a cocycle, and $\langle \phi - \phi'\rangle \equiv \left((\phi - \phi' + \left\lfloor{N/2}\right\rfloor) \mod N\right) - \left\lfloor{N/2}\right\rfloor$. This choice of cocycle is a slight modification of the more common choice~\cite{Gerbs}, where no addition or substraction of $\left\lfloor{N/2}\right\rfloor$ is required.  

A technical advantage of this branching structure is that any $A$ sublattice point, say $r_a$, appears first in the ordering of the vertices in all the triangles surrounding it. 
Consequently the ratio of $A_{ \{ \phi_r\} + 1_{r_a}}$ over $A_{\{\phi_r\}}$, where $\{ \phi_r\} + 1_{r_a}$ is the set $\{ \phi'_r \}$ with $\phi'_{r} = \phi_{r} + \delta_{r,r_a} \mod N$, depends only on the points surrounding $r_a$ (indexed in a clockwise fashion by $r_{a,k}$ with $k=1,\dots,6$):
\begin{align}
\label{Eq:Ratio}
\frac{A_{ \{ \phi_r\} + 1_{\phi_a}}}{A_{\{ \phi_r\}}} &= e^{\frac{2\pi i p}{N} {\rm d}_{2,r_a}}, \\  
 {\rm d}_{2,r_a} &= \frac{\sum_{k=1}^6 (-1)^k \langle (-1)^k (\phi_{r_{a,k+1}} - \phi_{r_{a,k}}) \rangle}{N}. 
\end{align}
For ``smooth'' configurations, where $|\left\langle \phi - \phi' \right\rangle| \ll N$ for neighboring sites, the function ${\rm d}_{2,r_a} \in \mathbb{Z}$ is a discrete analog of the notion of vorticity ~\cite{Ringel2015}.
Since $\sum_{r_a}  {\rm d}_{2,r_a} = 0$ on the torus, we obtained an extra global $Z_N$ symmetry, corresponding to rotating $\phi$ only on the $A$ sublattice.

The partition function $Z$ can now be simplified considerably by summing out all the variables on the $A$ sublattice ($\phi_{r_a}$). 
For $N$ and $p$ coprime, the cancellation of phases of the exponential in Eq. (\ref{Eq:Ratio}) leaves only the subset of configurations on the $B$ and $C$ sublattices $\{ \phi_{r_{b,c}}\}_0$ for which ${\rm d}_{2,r_a} \mod N = 0$ on every $A$ sublattice point.
This yields 
\begin{equation}
\begin{aligned}
Z &= \sum_{\{ \phi_{r_{b,c}}\}} \sum_{\{ \phi_{r_a}\} } A_{ \{ \phi_r\}} = N^{\#A} \sum_{{\{ \phi_{r_{b,c}}\}}_0} 1
\end{aligned}
\end{equation}
where $\#A$ is the number of $A$ sublattice sites and we also used the cocycle property that $\nu_m(0,0,\phi_{2},\phi_{3})=1 \ \forall \ \phi_{2},\phi_{3}$. 
Notably, we arrived at a classical partition function with real positive Boltzmann weights and thus any CFT that might describe local observables in this theory should be unitary.
 
To obtain the CFT associated with $Z$, we follow a heuristic approach which works for spin-ice models wherein a discrete, lattice zero-divergence constraint is promoted to an equation of motion for a continuous field \cite{Henley2005}. 
The analog of the spins in the spin-ice model is given by the link variables $\langle \phi_i-\phi_j \rangle$ of our model.
A slight difference is that ${\rm d}_2(r_a) = 0$ is a zero-curl constraint for these link variables rather than a zero-divergence constraint. 
Consider promoting $\phi$ to a compact boson $\varphi$ using the embedding $\varphi = 2\pi \phi/N \in [0,2\pi)$, a possible candidate for the Lagrangian is $\mathcal{L}=\frac{g}{4\pi} (\nabla \varphi)^2$. 
Using a duality transformation, the Lagrangian becomes $\mathcal{L}=\frac{1}{4\pi g} (\nabla \theta)^2$, with $\theta$ being the dual field of $\varphi$ such that $\nabla^2 \theta$ is the vortex density of $\varphi$ \footnote{See C. L. Kane's Lectures on Bosonization http://www.physics.upenn.edu/~kane/pedagogical/boulderlec12.pdf}. 
The equation of motion of $\theta$ is given by $\nabla^2 \theta = 0$ and can be seen as a continuum version of the zero-curl constraint we have found on the lattice. 

Based on the above approach, a sensible continuum action is given by
\begin{equation}
\begin{aligned}
S &= \int d^2 r \frac{g}{4\pi} (\partial \varphi)^2 + \lambda (V_{e=N} + V_{e=-N}) \\ 
&+ \lambda' (V_{m=N} + V_{m=-N})
\label{action}
\end{aligned}
\end{equation}
where $g$ is the stiffness, $V_{e,m}$ is the vertex operator of electric charge $e$ and magnetic charge $m$, with scaling dimension $\Delta_{e,m} = e^2/2g + gm^2/2$, and where only the most relevant vertex terms allowed by symmetry were kept.
In terms of $\varphi$ and its dual variable $\theta$, the electric and magnetic operators are given by $V_e = e^{i e \varphi}$ and $V_m = e^{i m \theta}$.
%
Considering ``smooth" configurations of $\phi$, the microscopic ${\rm d}_{2,r_a}=0$ condition translates under the embedding to allowing only magnetic charges which are multiples of $N$, i.e. $m\in N \mathbb{Z}$, in the action.
Likewise, the discrete nature of $\phi$ restricts the electric charges allowed in the action to respect the $Z_N$ subgroup of $U(1)$, i.e. $e \in N \mathbb{Z}$.
If both these magnetic and electric terms are non-relevant, which is true for $N^2/4 \geq g \geq 4/N^2$, the theory is attracted to a Gaussian fixed point with $c=1$ ~\cite{DiFran}.

To associate CFT primary operators with the $O_{\pm}(r)$ appearing in Eq.~\ref{Eq:SPTCFT}, we require two reasonable assumptions in addition to the above embedding: (1) The CFT operators should transform according to a given representation of the sublattice symmetry group. In the present case, after having integrated out the $A$ sublattice, the only remaining sublattice symmetry corresponds to an interchange of the $B$ and $C$ sublattices. (2) The CFT fusion rules should be consistent with the $Z_N$ additive relation for the charges. This is made possible by the presence of the vertex terms in the action, which corresponds to a finite density of screening charges with $e,m \in N\mathbb{Z}$.

\begin{figure}[t]
\includegraphics[width=50mm, clip=true]{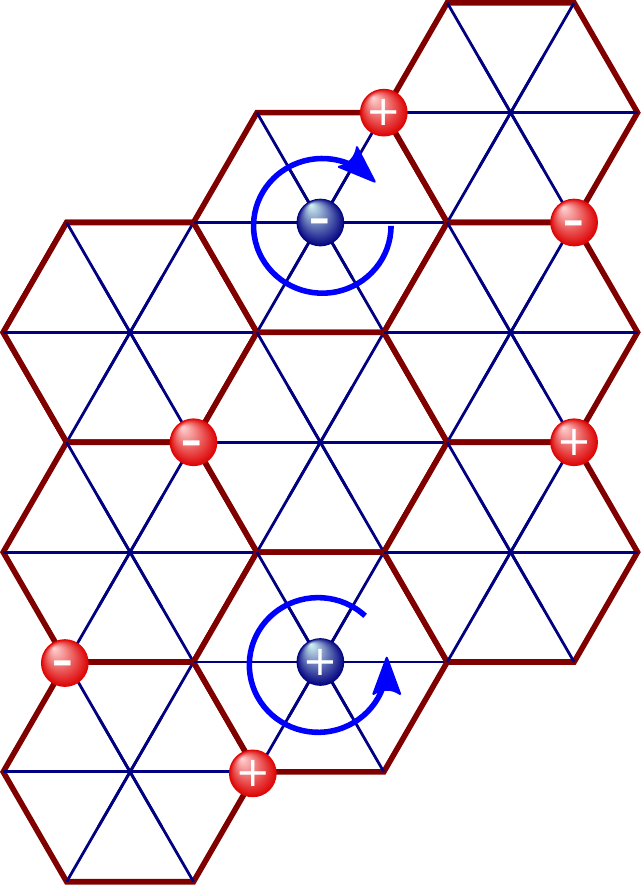}
\caption{Electric particles live on the $B$ and $C$ sublattices (shown in red) and magnetic particles live on the $A$ sublattice (shown in blue). The blue curvy arrows indicate the magnetic-particles-generated flux that is seen by the electric particles.}
\label{Fig:ElectricAndMagnetic}
\end{figure}

For $O_{\pm}(r)$ operators on the $B$ or $C$ sublattice, the embedding implies $e^{\mp i 2\pi \phi(r)/N} \rightarrow e^{\mp i \varphi(r)} $, yielding $O_{\pm} \rightarrow V_{e=\mp1}$.
Numerically (see Appendix \ref{Sec:NumericalBosons}), we find that there is a sign difference between $B$ and $C$: on the $B$ ($C$) sublattice, we have $O_{\pm} \rightarrow V_{e=\mp1}$ ($O_{\pm} \rightarrow -V_{e=\mp1}$), implying that $V_{e=\mp1}$ picks up a minus sign under the sublattice symmetry, as allowed by assumption (1).
For a charge $\alpha$ on the $A$ sublattice point $r_a$, the summation over $\phi_{r_a}$ enforces a non-zero vorticity ${\rm d}_{2,r_a}$ whose value is determined by $p {\rm d}_{2,r_a} - \alpha = 0 \mod N $.
The embedding thus associates a non-zero discrete vorticity ${\rm d}_{2,r_a}$ for $\phi$ to a magnetic charge with $m={\rm d}_{2,r_a}$ for $\varphi$.
For $p=1$, this simply leads to $O_{\pm} \rightarrow V_{m=\pm 1}$.
For $p>1$, the discussion remains the same if, on the $A$ sublattice, one enumerates the charge basis by the position of charge $\pm p$ particles instead of charge $\pm 1$ particles, since one has $O_{\pm}^p \rightarrow V_{m=\pm 1}$.
Different values of $p$ therefore simply correspond to a reshuffling of the different charge values, and we will use $p=1$ in the following in order to simplify notations.
In short, charges on the $A$ sublattice correspond to magnetic charges and charges on the $B$ and $C$ sublattices correspond to electric charges (see Fig.~\ref{Fig:ElectricAndMagnetic}).
The case $N=2$, made special by the fact that $O_{+} \equiv O_{-}$, is discussed in Appendix \ref{Sec:NumericalBosons}.

In Appendix \ref{Sec:NumericalBosons}, we numerically confirmed the above predictions for $N=2$ and $N=3$ in terms of central charge, operator content, power law behavior of correlation functions and representation of lattice operators.
We find $g=1\pm 5\e{-4}$ for $N=2$ and $g =0.925 \pm 0.01$ for $N=3$. 
Note that the value of $g$ is in general not universal, except possibly when an extra symmetry is present, as is the case for $N=2$ for which adding 1 to all the $\phi_r$ on any given sublattice is a symmetry.
This symmetry between the three sublattices corresponds to a symmetry between electric and magnetic charges in the CFT language and thereby forces $g$ to be at its electric-magnetic dual value of 1~\cite{Ginsparg}.

It is now illuminating to study the symmetry action at the light of the previous results.
The global symmetry action corresponds to applying $\phi \rightarrow \phi+ 1$ on each site.
At the CFT level, this symmetry action translates into different transformations depending on the sublattice.
On $B$ and $C$, the symmetry action translates via the embedding to $\varphi \rightarrow \varphi + \frac{2 \pi}{N}$.
On the $A$ sublattice, inserting the microscopic operator $e^{ \frac{-2 \pi i p \phi}{N}}$ corresponds to adding a vorticity of one, or equivalently to inserting a $V_{m=1} \equiv e^{i \theta}$ operator in the CFT language. 
It is then natural to associate the microscopic variable $-2\pi p \phi /N$ on the $A$ sublattice with $\theta$ in the CFT and, consequently, to associate the microscopic symmetry action $\phi \rightarrow \phi+1$ to $\theta \rightarrow \theta - \frac{2 \pi p}{N}$.
Thus, the global $Z_{N>2}$ symmetry, for non-zero $p$, rotates both $\theta$ and $\varphi$ simultaneously. 
The above symmetry action is exactly the one discussed in Ref.~\onlinecite{YuanMing2012} with regards to the edge theory of such SPTs. 
As shown in that work, given such an action of the symmetry, the only way a symmetry-preserving term can gap out the theory is by inducing a spontaneous symmetry breaking. 


We now go back to the microscopic derivation of the wavefunction amplitude in the symmetry-charge basis, $A_{ \{ \pm_i,r_i\}}$.
Focusing on $N>2$, we remove fast sublattice oscillations by defining $A_{ \{ e_i,m_i,r_i\}} \equiv \prod_i (-1)^{\delta_{r_i \in C}} A_{ \{ \pm_i,r_i\}}$, where $\delta_{r_i \in C}$ is 1 (0) if $r_i$ is on the $C$ ($A,B$) sublattice, and the set of electric and magnetic charges  ($\{ e_i,m_i,r_i\}$) is determined from $\{ \pm,r_i\}$ by the previous association (e.g. $(+,r_a) \rightarrow (0,+1,r_a)$).
The wavefunction amplitude can now be written as $A_{ \{ e_i,m_i,r_i\}} = \left\langle  \prod_i V_{e_i,m_i}(r_i) \right\rangle$, where the average value is taken with the action given by Eq.~\ref{action}.
The usual renormalization group procedure can be used from the lattice scale $a_0$ to the scale of average interparticle distance $l$ ($\rho=a_0^2/l^2$), leading to a suppression of $\lambda$ ($\lambda'$) as $\sim (a_0/l)^{\Delta-2}$ with $\Delta=N^2/2g$ ($\Delta=gN^2/2$).
We focus first on the infinitely dilute charge limit $\rho\to0$, for which $\lambda,\lambda'\to0$, yielding a simple form for the wavefunction amplitude:
\be
\bbea
A_{ \{ e_i,m_i,r_i\}} &= \prod_{k<l \in \mathcal{E}} |z_k - z_l|^{e_k e_l/g} \prod_{s<t \in \mathcal{M}} |z_s - z_t|^{m_s m_t g} \\ 
  &\prod_{k \in \mathcal{E}, t \in \mathcal{M}} \left(\frac{z_{k} - z_{t}}{|z_{k} - z_{t}|}\right)^{e_k m_t}
\eeea
\label{Eq:LaughlinLike}
\ee
where $z \equiv r_x + i \ r_y$ and where we used the fact that each particle is either electric or magnetic, but not both, to separate the particle indices into two subsets, $\mathcal{E}$ and $\mathcal{M}$, for electric and magnetic charges, respectively.
Notably, unless $\sum_i e_i = \sum_i m_i = 0$, the above correlator vanishes. 
Consequently, we find an enhanced $U(1) \times U(1)$ symmetry compared to the original wavefunction for which the above neutrality condition was only obeyed modulo $N$.

At small but finite $\rho$, $\lambda$ and $\lambda'$ are non-zero and act as a condensate of charge $\pm N$ particles thereby breaking the $U(1) \times U(1)$ symmetry present at $\rho=0$ down to the initial $Z_N \times Z_N$ symmetry. This mechanism is explained in Appendix \ref{Sec:DustyLaughlin}, where we also give a self consistent argument showing that the Laughlin picture persists even with a finite condensate density.  

Interestingly, in Ref.~\cite{Senthil2013} a similar $U(1) \times U(1)$ wavefunction was proposed (however with only positive charges). Consistently with our picture, condensing charge $N$ particles at the level of the Chern Simons theory associated with that wavefunction was shown to yield the $N$ different $Z_N$ SPT phases~\cite{YuanMing2012}.

\subsubsection{Type $\rn{2}$ cocycles}
Let $\Phi = (\phi_{\RN{1}},\phi_{\RN{2}})$ denote an element of $Z_N \times Z_N$, with $N$ prime. 
In the following, we will use $\eta=\RN{1},\RN{2}$ as the index that distinguishes between the two groups.
According to Ref. (\onlinecite{deWild}), the third cohomology group of $Z_N \times Z_N$ is spanned by products of two distinct types of 3-cocycles. 
The type $\rn{1}$ cocycles involve just one of the $Z_N$ groups while type $\rn{2}$ cocycles involve both. 
Namely,  
\begin{equation}
\begin{aligned}
\label{CocycleBis}
\nu_{\rn{1},p_{\eta}}&(0,\Phi_1,\Phi_2,\Phi_3) = e^{\frac{ 2 \pi i p_{\eta} \phi_{1,\eta} f_{\eta}}{N}}, \\ 
\nu_{\rn{2},p'}&(0,\Phi_1,\Phi_2,\Phi_3) = e^{\frac{ 2 \pi i p' \phi_{1,\RN{1}} f_{\RN{2}}}{N}}, \\ 
f_{\eta} &= \frac{\langle \phi_{2,\eta} - \phi_{1,\eta} \rangle + \langle \phi_{3,\eta} - \phi_{2,\eta} \rangle - \langle \phi_{3,\eta}-\phi_{1,\eta} \rangle}{N}
\end{aligned}
\end{equation}
where we again choose the branching structure given in Fig.~\ref{Fig:BranchingStructure} such that, for each triangle, $\Phi_1$ (resp. $\Phi_2$, $\Phi_3$) refers to the site on the $A$ (resp. $B$, $C$) sublattice.
A generic cocycle is then labelled by $(p_{\RN{1}},p_{\RN{2}},p')$ and given by $\nu_{\rn{1},p_{\RN{1}}}\nu_{\rn{1},p_{\RN{2}}}\nu_{\rn{2},p'}$. 

\begin{table}
\caption{\label{AbelianCases}Association of CFT primaries and microscopic operators for several diffent cocycles of $G = Z_N \times Z_N$. Regarding microscopic operators, $\phi_{a,\eta}$ stands for the insertion of an operator $e^{i \frac{2\pi}{N}\phi_{\eta}}$ on the $A$ sublattice and $\phi_{b(c),\eta}$ stands for the insertion of an operator $e^{i \frac{2\pi}{N}\phi_{\eta}}$ on the $B$ or $C$ sublattice. The CFT operator $V_{e(m),\eta}$ is an electric (magnetic) operator for the compactified boson indexed by $\eta$.
} 
\begin{ruledtabular}
\begin{tabular}{c|c|c|c|c|c|c|c}
 $p_{\RN{1}}$ &$p_{\RN{2}}$& $p'$& $V_{m,\RN{1}}$ & $V_{e,\RN{1}}$ & $V_{m,\RN{2}}$ & $V_{e,\RN{2}}$ & Central charge \\
\hline
 $\neq 0$ & $0$ & $0$ & $\phi_{a,\RN{1}}$ & $\phi_{b(c),\RN{1}}$& & & 1 \\
$\neq 0$ & $\neq 0$ & $0$ &$\phi_{a,\RN{1}}$ & $\phi_{b(c),\RN{1}}$ &  $\phi_{a,\RN{2}}$ & $\phi_{b(c),\RN{2}}$ & 2 \\
 $0$ & $0$ & $\neq 0$ &  & &  $\phi_{a,\RN{1}}$ & $\phi_{b(c),\RN{2}}$  & 1 \\
 $0$ & $\neq 0$ & $\neq 0$ &  & &  $\phi_{a,\RN{1}}+\phi_{a,\RN{2}}$ & $\phi_{b(c),\RN{2}}$  & 1
\end{tabular}
\end{ruledtabular}
\end{table}


We consider first some simple cases, as summarised in Table \ref{AbelianCases}.
First, when the only non-trivial cocycle is a type $\rn{1}$ cocycle for one of the two groups, say the first one, the previous arguments for $G=Z_N$ can be applied to the first group, while the degrees of freedom of the second group form a trivial paramagnet. Second, when there are two non-trivial type $\rn{1}$ cocycle, one for each group, one finds two decoupled $G=Z_N$ SPT phases, and again the previous arguments apply directly. 
Third, when $p' \neq 0$ but $p_{\RN{1}} = p_{\RN{2}} = 0$, one should sum $\phi_{\RN{1}}$ on the $A$ sublattice and obtain a compactified boson ($\varphi_{\RN{2}}$) associated with the coarse graining of $\phi_{\RN{2}}$ degrees of freedom on the $B$ and $C$ sublattices. 
For $N>2$, the electric charge operators $V_{e=\pm1}$ of that boson would correspond to the microscopic operators $e^{\frac{\pm 2 \pi i \phi_{\RN{2}}}{N}}$ on the $B$ and $C$ sublattice as before.
However, the magnetic charge operators $V_{m=\pm1}$ would now correspond to $e^{\frac{\pm p' 2 \pi i \phi_I}{N}}$ on the $A$ sublattice.
All the other microscopic degrees of freedom ($\phi_{\RN{1}}$ on the $B$ and $C$ sublattice and $\phi_{\RN{2}}$ on the $A$ sublattice) form a trivial paramagnet.
Lastly, for $p' \neq 0$ and $p_{\RN{2}} \neq 0$ but $p_{\RN{1}} = 0$, one obtains a single compactified boson $\varphi_{\RN{2}}$ associated with $\phi_{\RN{2}}$. For $N>2$, the microscopic operators $e^{\frac{\pm 2 \pi i \phi_{\RN{2}}}{N}}$ on $B$ and $C$ correspond to $V_{e=\pm1}$ and the microscoic operator $e^{\frac{\pm 2 \pi i (p_{\RN{2}}\phi_{\RN{2}} + p' \phi_{\RN{1}})}{N}}$ on the $A$ sublattice corresponds to $V_{m=\pm 1}$.

 
A more complicated case, which does not follow directly from the previous $G=Z_N$ results, is $p' \neq 0$ and $p_{\RN{1}} \neq 0$ but $p_{\RN{2}} = 0$. 
In this case, one can integrate out $\phi_{\RN{1}}$ on the $A$ sublattice to obtain that the condition $(p_{\RN{1}} {\rm d}_{2,r_a}[\phi_{\RN{1}}] + p' {\rm d}_{2,r_a}[\phi_{\RN{2}}]) \mod N = 0 $ has to be satisfied for each $B,C$ hexagon centered at $r_a$.
The discrete vorticity ${\rm d}_{2,r_a}[\phi]$ is defined in Eq.~\ref{Eq:Ratio}. It is then natural to consider the linear transformation $(\tilde\phi_{\RN{1}},\tilde\phi_{\RN{2}}) = (p_{\RN{1}} \phi_{\RN{1}},p_{\RN{1}} \phi_{\RN{2}}) + (p' \phi_{\RN{2}},-p' \phi_{\RN{1}})$, defined in the two-dimensional vector space over the finite field $\mathbb{F}_N$ (which is well-defined since $N$ is prime).
This linear transformation is an isomorphism if its determinant is non-zero, i.e. if $(p_{\RN{1}}^2 + p'^2) \mod N \neq 0$.
We now focus on the case $N\gg 1$ and $p_{\RN{1}},p' \gtrsim 1$ for which this condition is fulfilled.

One can then repeat the assumption made previously that the important configurations are those in which $\phi_{\RN{1}}$ and $\phi_{\RN{2}}$ change by much less than $N/2$ between adjacent lattice sites. 
In this case, one has that $p_{\RN{1}} {\rm d}_{2,r_a}[\phi_{\RN{1}}] + p' {\rm d}_{2,r_a}[\phi_{\RN{2}}] = {\rm d}_{2,r_a}[\tilde\phi_{\RN{1}}] $ and a compactified boson $\tilde\varphi_{\RN{1}}$ emerges, corresponding to the coarse graining of $\tilde\phi_{\RN{1}}$. 
Repeating the rationale used for $G=Z_N$, one would then find that the microscopic operator $e^{\frac{\pm 2 \pi i \phi_{\RN{1}}}{N}}$ on the $A$ sublattice corresponds to $V_{m=\pm1}$ and that the microscopic operator $e^{\frac{\pm 2 \pi i \tilde\phi_{\RN{1}}}{N}}$ on the $B$ or $C$ sublattice corresponds to $V_{e=\pm 1}$.
All the other microscopic degrees of freedom ($\phi_{\RN{2}}$ on $A$ and $\tilde\phi_{\RN{2}}$ on $B$ and $C$) form a trivial paramagnet. A test of this conjecture, as well as a generalization for all $(p_{\RN{1}},p_{\RN{2}},p')$ and non-prime $N$, is left for future work. 


\subsubsection{Type 3 cocycle and exact mapping to loop models}
When a product of more than two $Z_N$ groups is considered, the third cohomology group is spanned by the type $\rn{1}$ and $\rn{2}$ 3-cocycles considered above, as well as one additional distinct 3-cocycle, called type $\rn{3}$ \cite{deWild}. 
Focusing on the case of $G=Z_N \times Z_N \times Z_N$ and denoting group elements as $\Phi=(\phi_{\RN{1}},\phi_{\RN{2}},\phi_{\RN{3}})$, the type $\rn{3}$ 3-cocycle is given by
\begin{align}
\label{Z2Z2Z2Cocycle}
\nu_{\rn{3},p}(0,\Phi_1,\Phi_2,\Phi_3) &= e^{\frac{ 2 \pi i p a_{\RN{1}} b_{\RN{2}} c_{\RN{3}} }{N}} \\ \nonumber 
a_{\RN{1}} &= \phi_{1,\RN{1}} \\ \nonumber 
b_{\RN{2}} &= \langle \phi_{2,\RN{2}} - \phi_{1,\RN{2}} \rangle \\ \nonumber 
c_{\RN{3}} &= \langle \phi_{3,\RN{3}} - \phi_{2,\RN{3}} \rangle
\end{align} 
where we again choose the branching structure given in Fig.~\ref{Fig:BranchingStructure} such that, for each triangle, $\Phi_1$ (resp. $\Phi_2$, $\Phi_3$) refers to the site on the $A$ (resp. $B$, $C$) sublattice.
Notably this cocycle is not defined in terms of the $f$ factor present in the previous examples (Eqs. \ref{Cocycle} and \ref{CocycleBis}) and the reasoning used for $G=Z_N$ appears irrelevant in this case. 

Instead, consider re-writing this 3-cocycle as 
\begin{align}
\label{Z2Z2Z2}
\nu_{\rn{3},p}(0,\Phi_1,\Phi_2,\Phi_3) &= \left[\tilde{\nu}_p(0,\tilde{\Phi}_1,\tilde{\Phi}_2)\right]^{c_{\RN{3}}}
\end{align}
where $\tilde{\Phi}_i = (\phi_{i,\RN{1}}, \phi_{i,\RN{2}})$ and where $\tilde{\nu}_p$ can be recognized as a 2-cocycle associated with a $G= Z_N \times Z_N$ symmetry \cite{Ringel2015} (i.e. a cocycle in the second cohomology group ${\mathcal H}^2(Z_N \times Z_N,U(1))$).

This suggests that type $\rn{3}$ 3-cocycles are described by the decorated domain wall picture of SPTs \cite{AshvinDecorated}. 
Following this picture, each domain wall for the $\phi_{\RN{3}}$ degrees of freedom should be thought of as carrying its own $G=Z_N \times Z_N$ 1D-SPT phase for the $\phi_{\RN{1}}$ and $\phi_{\RN{2}}$ degrees of freedom.
[As a technical comment, note that the 2-cocycles of the form given by $\tilde\nu_p$ span ${\mathcal H^2}(Z_N \times Z_N,{\mathcal H^1}(Z_N,U(1)))$, which is the subgroup of ${\mathcal H^3}(Z_N \times Z_N \times Z_N,U(1))$ for which the decorated domain wall picture applies \cite{AshvinDecorated}].  

\begin{figure}[t]
\includegraphics[width=50mm, clip=true]{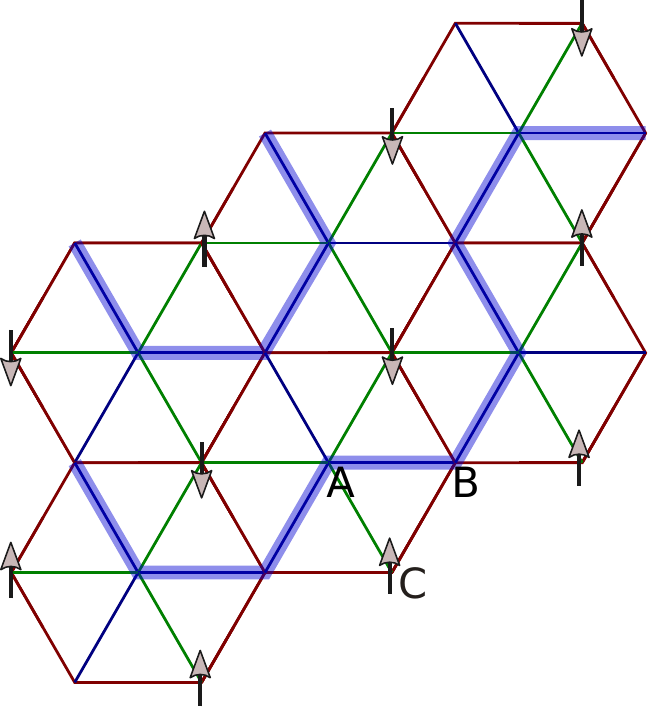}
\caption{Decorated domain walls picture emerging naturally for our choice of cocycles and branching structure for the case of $G=Z_2 \times Z_2 \times Z_2$. The arrows give the value of $\phi_{\RN{3}}(r_c)$ (the restriction of $\phi_{\RN{3}}$ to the $C$ sublattice) and the shaded blue lines give the domain walls of the $\phi_{\RN{3}}(r_c)$ configuration, which form non-intersecting loops on the hexagonal lattice formed by the $A$ and $B$ sites. $AB$ edges are shown in blue, $BC$ edges in red and $AC$ edges in green.
}
\label{Fig:Z2Z2Z2}
\end{figure}

{\bf The $N=2$ case:} For $N=2$, the only non-trivial case corresponds to $p=1$. Furthermore, since the cocycle is real in this case, one has $\nu_{\rn{3},1} = \nu_{\rn{3},1}^{-1}$ and one can ignore the orientation of the triangles. First, let us consider the restriction of $\phi_{\RN{3}}(r)$ to its values on the $C$ sublattice only, $\phi_{\RN{3}}(r_c)$.
Since the $C$ sublattice is itself a triangular lattice, its dual lattice is an hexagonal lattice, and it is the one formed by the $A$ and $B$ sublattices.
As shown in Fig.~\ref{Fig:Z2Z2Z2}, this means the domain walls of $\phi_{\RN{3}}(r_c)$ form non-intersecting loops on the $A,B$ hexagonal lattice.
We now show that these loops carry 1D SPTs for $(\phi_{\RN{1}}, \phi_{\RN{2}})$.

From Eq.~\ref{Z2Z2Z2}, it is clear that the only triangles with a non-trivial factor are the ones for which $c_{\RN{3}} = 1$, i.e. the ones for which $\phi_{\RN{3}}$ is flipped when going from the $B$ site to the $C$ site of the triangle.
If this is the case, we call the triangle active and its non-trivial factor is given by the 2-cocycle $\tilde\nu_p$ for the $(\phi_{\RN{1}}, \phi_{\RN{2}})$ degrees of freedom living on the two sites of the $AB$ edge of the triangle.
Now, since each $AB$ edge (shown in blue in Fig.~\ref{Fig:Z2Z2Z2}) is shared by two triangles, there are two possibilities: either (1) the $AB$ edge is occupied by a $\phi_{\RN{3}}(r_c)$ domain wall (highlighted in shaded blue in Fig.~\ref{Fig:Z2Z2Z2}), in which case the $C$ sites of the two triangles have a different value for $\phi_{\RN{3}}$, or (2) the $AB$ edge is unoccupied by a $\phi_{\RN{3}}(r_c)$ domain wall, in which case the $C$ sites of the two triangles have the same value for $\phi_{\RN{3}}$.
Crucially, for the first possibility, we know that one of the two triangles is active and the other one is inactive, since, when going from one $C$ site to the other along the $BC$ edges of these two triangles, there has to be exactly one $\phi_{\RN{3}}$ flip.
In this case, one obtains a factor of $\tilde\nu_p$ for the $AB$ edge.
For the second possibility, however, we know that the two triangles are both inactive or both active, since, when going from one $C$ site to the other along the $BC$ edges of these triangles, there has to be, respectively, zero or two $\phi_{\RN{3}}$ flips.
In this case, the overall factor is trivial, since $\tilde\nu_p^2=1$.

Now that we have shown that a $\tilde\nu_p$ factor appears for each $AB$ edge covered by a $\phi_{\RN{3}}(r_c)$ domain wall, we turn our attention to the $(\phi_{\RN{1}}, \phi_{\RN{2}})$ degrees of freedom along one such domain wall.
Since the vertices along this domain wall alternate between the $A$ and $B$ sublattice, it is advantageous to split the domain wall as a succession of $AB$ pairs indexed by $i$ and for which the two vertices inside the pair are indexed by $u=A,B$. This leads to the following notation for the degrees of freedom along one domain wall: $(\phi_{i,u,\RN{1}}, \phi_{i,u,\RN{2}})$.
The $\tilde{\nu}_p$ factors then generate the following overall phase factor for the entire domain wall:
\begin{align}
&e^{ i \pi \sum_i \left[ \phi_{i,A,\RN{1}} (\langle \phi_{i,B,\RN{2}}-\phi_{i,A,\RN{2}} \rangle + \langle \phi_{i-1,B,\RN{2}} -\phi_{i,A,\RN{2}} \rangle) \right] }  \\ \nonumber 
&= e^{ i \pi \sum_i \left[ \phi_{i,A,\RN{1}} (\phi_{i,B,\RN{2}}- \phi_{i-1,B,\RN{2}}) \right] }  \\ \nonumber 
&= e^{ i \pi \sum_i \left[ \phi_{i,B,\RN{2}} (\phi_{i,A,\RN{1}}- \phi_{i+1,A,\RN{1}}) \right] } 
\end{align}
where we have used the $Z_2$ algebra obeyed by $\phi$ and the last line follows from discrete integration by parts (and assuming a closed domain wall, which is always the case for periodic boundary conditions). 
As shown elsewhere~\cite{AshvinDecorated,Ringel2015}, these phase factors describe a 1D SPT ground state with $G=Z_2 \times Z_2$. Furthermore, correlations in the 1D Stat. Mech. model defined by these amplitudes are long ranged \cite{Cenke2014}. Indeed, one may integrate out all $\phi_{i,A,\RN{1}}$ (resp. $\phi_{i,B,\RN{2}}$), and obtain that all $\phi_{i,B,\RN{2}}$ (resp. $\phi_{i,A,\RN{1}}$) must be equal, and hence are long-range correlated. 

Collecting these results, we find that there is a two to one mapping between $\phi_{\RN{3}}$ configurations and loops (i.e. domain walls) on the hexagonal $AB$ lattice.  
For a given $\phi_{\RN{3}}$ configuration, the correlation for the remaining degrees of freedom $(\phi_{\RN{1}}, \phi_{\RN{2}})$ at different sites is strictly zero unless these sites lie on the same $\phi_{\RN{3}}(r_c)$ domain wall, in which case the correlation is infinitely-ranged along the domain wall.
Calculating a correlation of the type $C_{\RN{2}}(r_b,r'_b) \equiv \langle e^{i\pi \phi_{r_b,\RN{2}}} \ e^{i \pi \phi_{r'_b,\RN{2}}} \rangle$ therefore amounts to evaluating the probability of $r_b$ and $r'_b$ lying on the same $\phi_{\RN{3}}(r_c)$ domain wall. This is precisely the definition of the correlation function of 2-leg watermelon operators, as defined in Section \ref{secLG}. 

To determine the behavior of the above watermelon operator, we need to identify what is the precise loop model that was obtained. These are defined by two parameters, $n$ and $x$, denoting the loop fugacity and loop tension, respectively. Fixing $\phi_{\RN{3}}$, and summing up, say, $\phi_{\RN{1}}$, one obtains that $\phi_{\RN{2}}$ degrees of freedom along a loop are all identical. Since they can all either be $0 \in Z_2$ or $1 \in Z_2$, this gives a weight of $n=2$ for each loop. 
Loop tension comes from the fact that the $\phi_{\RN{2}}$ degrees of freedom that lie on loops cannot fluctuate freely. 
Thus the presence of a loop of length $l$ reduces the partition function by a factor of $2^{-l/2}$. 
The loop tension is therefore given by $x=1/\sqrt{2}$.
Interestingly, this loop model is right at an exactly solvable point of the honeycomb seven-vertex model \cite{Jacobsen} described by the $SU(2)_1$ Wess-Zumino-Witten (WZW) CFT.
In terms of the Coulomb gas description given in Section \ref{secLG}, this corresponds to having zero background charge ($e_0=0$), and this theory is therefore a free boson with $g=1$ and $c=1$.
The probability that two points at a distance $r$ are on the same domain wall is given by $r^{-2 \Delta_{l=1}}=r^{-1}$,  where $\Delta_l = -e_0^2/2g + g l^2/2$ gives the scaling dimension of the $2l$-leg watermelon operator. This exponent governs correlations functions of microscopic charge operators of the type $e^{i\pi \phi_{r_{a/b},\RN{1}/\RN{2}}}$, as exemplified by $C_{\RN{2}}(r_b,r'_b)$.

{\bf The $N=3$ case:} Considering the case of $N>2$, domains walls may now branch and intersect. Repeating the previous reasoning, one ends up with loop models in which the loops may branch. More accurately, each domain-wall carries a group element and the branchings must obey the loop algebra. Such loop models cannot be solved straightforwardly using the Coulomb gas approach and thus we resort to a numerical calculation of the transfer matrix for $N=3$.
We found again a critical model, with a central charge of $c=2\pm0.05$. 
The first excited state of the transfer matrix corresponds to a scaling dimension of $0.4 \pm 0.008$ and is 6-fold degenerate.
Given the abundance of compact bosons we have found so far, a natural guess is that the CFT obtained here is that of two compact bosons. 
However, if these are decoupled, it is impossible to reproduce the observed degeneracy of $6$ for the lowest excited state using just the two parameters $g_1,g_2$. 
Considering two coupled compact bosons \cite{Ryu2015}, the moduli-space (or the space of critical theories) is spanned by four parameters \cite{Ryu2015} ($[G_{11},G_{12},G_{22},B_{12}]$). 
This space includes the $SU(3)_1$ WZW model model at $[1,0.5,1,0.5]$ and so all these models can be thought of as exactly marginal deformation of this model. 
Note that we have already encountered similar deformations since the models obtained previously for $G=Z_N$ can be viewed as exactly marginal deformations of an $SU(2)_1$ WZW model. 
While the central charge and 6-fold degeneracy can be reproduced by choosing the four aforementioned parameters within the appropriate region of parameter space, our finite-size numerics do not allow us to make a definite statement about the identification of this theory with a two-component Luttinger liquid.

To sum up, the $G=[Z_{N=2}]^3$ type 3 SPT corresponds to an $SU(2)_1$ WZW theory with $c=1$, while the $G=[Z_{N=3}]^3$ type 3 SPT corresponds to a critical theory with $c=2$ that could be related to a marginal deformation of an  $SU(3)_1$ WZW theory.
This qualitative difference between $N=2$ and $N=3$ is highly unusual and seems to indicate that the usual Chern-Simons reasoning \cite{YuanMing2012} within which one embeds the finite group in a product of $U(1)$ group does not apply in a straightforward way to type 3 cocycles.
Interestingly, the non-linear sigma model construction \cite{PhysRevB.91.134404} embeds the finite group in $SU(2) \times SU(2)$ for $D=2$ spatial dimensions.
While $[Z_{N=2}]^3$ is a finite group of $SU(2) \times SU(2)$, $[Z_{N=3}]^3$ is not, and a qualitative difference between these two cases is therefore consistent with this picture.
It would be highly interesting to study what happens for $N>3$.

\subsection{Non abelian symmetries: $D_3$}
\label{subsec:D3}
The simplest non-Abelian group to consider is the Dihedral group of the triangle ($D_3$). It has six elements ($g=(\Phi,\theta)$) corresponding to three rotations ($\theta \in [0,1,2]$) and an inversion $\Phi \in [0,1]$. Due to the non-Abelian nature of the group, it makes sense to use multiplicative notation for the group operation. Multiplication is given by $(\Phi,\theta)(\Phi',\theta') = ( \langle \Phi + \Phi' \rangle_2,\langle (-1)^{\Phi'} \theta + \theta' \rangle_3)$, where $\left\langle \dots \right\rangle_N$ stands for modulo $N$. The cohomology group here is ${\mathcal H}^{2}(D_3,U(1)) = Z_6$ and so 6 distinct SPT phases exist (including the trivial phase). The cocycles can be written as (see 6.20 in Ref. \onlinecite{ModularData}) 
\begin{align}
&\omega_p((\Phi_1,\theta_1),(\Phi_2,\theta_2),(\Phi_3,\theta_3)) = \\ \nonumber 
&e^{\left\{ -\frac{2\pi i p}{9} \left[(-1)^{\Phi_2+\Phi_3}\theta_1[(-1)^{\Phi_3} \theta_2 + \theta_3 - \langle (-1)^{\Phi_3} \theta_2 + \theta_3 \rangle_3 ] + \frac{9}{2}\Phi_1 \Phi_2 \Phi_3\right] \right\}}
\\ \nonumber 
&=\nu_p\left(1,(\Phi_1,\theta_1),(\Phi_1,\theta_1)(\Phi_2,\theta_2),(\Phi_1,\theta_1)(\Phi_2,\theta_2)(\Phi_3,\theta_3)\right). 
\end{align}

For $p=3$, a certain simplification arises: the factor $[(-1)^{\Phi_3} \theta_2 + \theta_3 - \langle (-1)^{\Phi_3} \theta_2 + \theta_3 \rangle_3 ]$ is always a multiple of $3$ and so it is effectively zero inside the above exponent. 
Consequently, the cocycle becomes $e^{ -\pi i p \Phi_1 \Phi_2 \Phi_3 }$, which is no other than the type 3 cocycle for $G'=Z_2\times Z_2 \times Z_2$ defined in Eq.~\ref{Z2Z2Z2Cocycle}, after the following identification: on the $A$ (resp $B$, $C$) sublattice, $\Phi$ maps to $\phi_{\RN{1}}$ (resp $\phi_{\RN{2}}$, $\phi_{\RN{3}}$).
%
%
%
Thus, up to a multiplication by some trivial factor coming from the completely uncorrelated $\theta$ degrees of freedom, the partition function for $p=3$ and for the type $\rn{3}$ cocycle are identical. 
They are thus both described by a $c=1$, $g=1$ compact boson. 

For $p=1$, we turn again to numerics. The large size of the group reduces our ability to approach the scaling limit. Still, looking at a circumference of $L=6$, we find that the leading scaling dimensions agree very well with $c=1$, $g=1$ (see Table (\ref{Table:D3})). This suggests that for all values of $p$, we obtain a $c=1$, $g=1$ compact boson.  Notably, however, the scaling dimensions here do not show exact degeneracies on the lattice, as was the case for all the Abelian cases. We attribute this to the fact that, for $p=1$, even with the $ABC$ branching structure, the wavefunction does not have an enhanced $D_3 \times D_3$ symmetry.

\subsection{The curious case of the $Z_2$ Levin-Gu wavefunction}
\label{secLG}
A different type of wave function to consider is the Levin-Gu wavefunction~\cite{Levin2012} for the Ising ($G=Z_2$) SPT whose phase factors  are not directly given by group-cohomology cocycles.  
For clarity, we consider $\phi_r=0,1$ as denoting the ${\sigma}^z_r=+1,-1$ eigenvalues of a spin $1/2$. 
Correspondingly, the charge $\alpha_r=0,1$ denotes the ${\sigma}^x_r=+1,-1$ eigenvalues. 
The amplitude $A_{ \{\sigma^z_r \}}$ is simply given by $(-1)$ to the number of domain walls in the $\sigma^z_r$ configuration.
This model flows to a stable critical phase with central charge $c=-7$ described in terms of a Coulomb gas with background charge $e_0=2/3$ and stiffness $g=1-e_0$ ~\cite{Nienhuis1982,Cardy2000,Jacobsen2003,Jacobsen,Cenke2014}.
There are two types of primary operators in this theory: electric operators $V_{e}(r) \equiv e^{i e \varphi(r)}$ with $e \in \mathbb{Z}$ and with scaling dimension $\Delta_e = e(e-2e_0)/2g$ and $2l$-leg ``watermelon'' operators $W_{l}(r)$ with $l \in \mathbb{Z}$ and with scaling dimension $\Delta_l = -e_0^2/2g + g l^2/2$.
The latter correspond to imposing $2l$ legs of domain walls emanating or closing at a given point~\cite{Jacobsen}.

Numerically, we find a good agreement with Coulomb gas predictions, despite the fact that we are dealing with a logarithmic CFT~\cite{Flohr2003}.
As explained in Appendix \ref{Sec:NumericalLoops}, the same prescription as before is used to find a correspondence between CFT primary operators and the microscopic operator $O(r) = \sigma^z_r \equiv e^{i \pi \phi_r}$.
In this case, we find that these operators lead to some linear combination of $V_{e=1}$ and of the 2-leg watermelon operator. Interestingly, the 2-leg watermelon operator has the same scaling dimension as $V_{e=1}$ and the 4-leg watermelon operator has the same scaling dimension as $V_{e=0}$, consistently with a $Z_2$ charge interpretation. 
We comment that, following a suitable choice of basis, the wavefunction amplitude still appears as a single CFT correlator (see Appendix \ref{Sec:NumericalLoops}). 
%


The attentive readers will have noticed that this is the second non-trivial $G=Z_2$ SPT wavefunction we have discussed in this work.
The first one was the group cohomology wavefunction, and its bulk CFT was given by a free boson ($c=1$).
It is interesting to note that these two wavefunctions have the same entanglement spectrum\footnote{This was checked numerically.} but different bulk CFTs.
While we leave for future work the discussion of the possible implications of this different bulk CFT, we note that the local unitary transformation taking the Levin-Gu wavefunction to the group cohomology wavefunction can be written as $\prod_r \sigma^z_r \prod_{\left\langle r r' \right\rangle} (-1)^{\phi_r \phi_{r'}}$, and was shown to have a non-trivial action on weak SPT indices \cite{2015arXiv150802695M}.
Another perhaps relevant observation is the fact that the central charges of the two different CFTs differ by 8, which could be an indication that the two states differ by a bosonic $E_8 \times \overline{E_8}$ bosonic SPT (with central charge 8 and chiral central charge 0). The physical $Z_2$ symmetry would then have to act non-trivially on the $E_8 \times \overline{E_8}$ state in order to protect it.

\section{Hidden order}

As we showed in the previous section, the $Z_N \times Z_N$ wavefunction can be written as 
\be
\bbea
A_{ \{ e_i,m_i,r_i\}} &= \prod_{k<l \in \mathcal{E}} |z_k - z_l|^{e_k e_l/g} \prod_{s<t \in \mathcal{M}} |z_s - z_t|^{m_s m_t g} \\ 
  &\prod_{k \in \mathcal{E}, t \in \mathcal{M}} \left(\frac{z_{k} - z_{t}}{|z_{k} - z_{t}|}\right)^{e_k m_t}
\eeea
\label{Eq:LaughlinLike}
\ee

Using this formula, we can discuss the presence of hidden order.
Based on this Laughlin-like picture, one can readily unveil a composite-particle~\cite{Jain2007} or hidden order~\cite{Girvin1987} structure behind these SPTs. Following Ref.~\cite{Girvin1987}, we use a suitable flux attachment transformation ($U$) to remove complex phases from $| \psi \rangle$, so that the following wavefunction is obtained 
\be
U|\psi\rangle = | |\psi| \rangle \equiv \frac{|Z|}{\mathcal{N}} \sum_{\{ \pm_i,r_i\}} |A_{ \{ \pm_i,r_i\}}| \ket{\{ \pm_i,r_i\}}.
\ee
From Eq. (\ref{Eq:LaughlinLike}) we find that the resulting amplitudes $|A_{ \{ e_i,m_i,r_i\}}|$ are related to two decoupled (electric and magnetic) two-component plasmas via $|A_{ \{ e_i,m_i,r_i\}}| = e^{-F[\{ e_i,m_i,r_i\}]}$ with
\be
\bbea
F[\{ e_i,m_i,r_i\}] &= - \sum_{k<l \in \mathcal{E}} \frac{e_k e_l}{g} \log(|r_k - r_l|) \\ 
&- \sum_{s<t \in \mathcal{M}} m_s m_t g \log(|r_s - r_t|).
\label{plasma} 
\eeea
\ee
Considering the correlator $C(r) \equiv \langle  \psi | U^{\dagger} e^{-i 2\pi \phi(0)/N} e^{ i 2\pi \phi(r)/N} U | \psi  \rangle$, we obtain 
\be
\bbea
C(r) & = |r|^{-\alpha} \sum_{\{ \pm_i,r_i\}} e^{-\tilde{F}[\{ \pm_i,r_i\};(+,0),(+,r)]} \\
&\propto |r|^{-\alpha} \text{ for } |r| \to \infty
\label{ModCorrel}
\eeea
\ee
where $0$ and $r$ are on the $A$ ($B$ or $C$) sublattice, where $\alpha = g /2$ ($\alpha=1/2g$) and where $\tilde{F}[\{ \pm_i,r_i\};(+,0),(+,r)]$ is the energy of a plasma as defined in Eq.~\ref{plasma} but with charges of magnitude $\sqrt{2}$ and with two additional test particles of magnetic charge $\pm 1/\sqrt{2}$ (of electric charge $\mp 1/\sqrt{2}$, respectively) at positions $0$ and $r$.
In order to obtain the second line of Eq.~\ref{ModCorrel}, one simply has to assume that the magnetic (electric) plasma screens. The screening condition for such plasmas is $g<2$ ($g>1/2$)~\cite{Plasma1,Plasma2,Plasma3,Plasma4,Bonderson2011}. Thus, using flux attachment, we find that $|\psi \rangle$ possesses quasi-long-range hidden order associated with electric operators which break the $Z_N$ symmetry.   

The Laughlin-like picture put forward here can also be used to study the charge fractionalization appearing after the insertion of symmetry fluxes (see App. \ref{App:Flux}).


\section{Integrability} 
In this section, we discuss the relation between SPT wavefunctions coming from the group cohomology construction and integrable models.
We show that (1) in several cases, there is an exact mapping between the auxiliary Stat. Mech. model of these SPTs and an integrable model, (2) even when such a mapping could not be found, the numerically obtained finite-size spectrum exhibits
the exact degeneracies predicted by CFT and (3) the finite-size corrections to scaling dimensions show better convergence than what would generically be expected.

We have found three examples where the auxiliary Stat. Mech. models corresponding to a group cohomology SPT wavefunction can be exactly mapped to an integrable model.
The first example is the case of $G=Z_2\times Z_2 \times Z_2$ with the type 3 cocycle, for which we have established an exact mapping to a loop model with loop fugacity $n=2$ and loop tension $x=1/\sqrt{2}$, which happens to be an integrable point.
At these values, it is known that its scaling limit is given by a $c=1$, $g=1$ compact boson, or in other words a $SU(2)_1$ WZW theory \cite{Jacobsen}. 
The second example appears for $G=Z_2$ and is obtained in a similar way since the partition function of the case of $G=Z_2\times Z_2 \times Z_2$ is equal to the one of $G=Z_2$ times that of a trivial paramagnet.
The mapping between the two descriptions works in the following way: on the $A$ (resp $B$, $C$) sublattice, $\phi$ maps to $\phi_{\RN{1}}$ (resp $\phi_{\RN{2}}$, $\phi_{\RN{3}}$).
The third example is that of $D_3$ discussed in Sec. \ref{subsec:D3}, which again maps to the $G=Z_2 \times Z_2 \times Z_2$ case.

Considering the case of $G=Z_3$, it cannot be mapped to the same $n=2$ loop model since the numerically obtained value of $g$ differs from $1$. 
Still, the numerical results point to some fine-tuned point: First, the numerically obtained finite-size spectra have the exact degeneracies predicted by CFT (see Table \ref{Table:Z3}) and, second, finite size corrections seem to be qualitatively smaller than generically expected.
Since the leading correction to scaling dimensions goes like $\mathcal{O}(1/L^{\Delta-2})$, where $\Delta$ is the scaling dimension of the most relevant operator allowed in the action, a microscopic model generically exhibits corrections that scale like $O(1/L^2)$ because of the operator $L_{-2}\bar{L}_{-2} \mathbbm{1}$ ($L_n$/$\bar{L}_n$ is the left/right Virasoro generator and $\mathbbm{1}$ is the identity operator) \cite{Cardy1986}.
If the lattice is triangular, non-conformal-scalar terms such as $L^{-3} \mathbbm{1}$ are also generically allowed in the action\cite{Cardy1986}, and would lead to $\mathcal{O}(1/L)$ corrections to scaling.
Instead, from the numerics carried out for $G=Z_3$, corrections to the scaling of the identity operator (a quantity which does not depend on our estimate of $g$) are given by $3 \times 10^{-3}$ for $L=10$, which seems to indicate that $O(1/L)$ corrections are fine tuned close to zero, and that even the usual $O(1/L^2)$ are reasonably small.
Similarly small corrections are present for all other scaling dimensions. 

This is in contrast to the case of the Levin-Gu wavefunction, for which the loop model based on the original wavefunction is given by $n=-1$, $x=1$, which itself is not an integrable point, but which is in the basin of attraction of the integrable point $n=-1$, $x= 1/\sqrt{2-\sqrt{3}}$.
There, numerical results obtained for the original wavefunction show criticality but almost no similarity with the conformal theory predictions even for the largest system size we reached ($L=24$).
On the contrary, the numerics performed at the integrable point show extremeley good convergence to the CFT predictions, as explained in Sec.~\ref{secLG}.

 \section{Conclusion}
In conclusion, we have proposed a CFT-based method to analyze SPT phases which reveals their properties in a transparent way and we showed how this description emerges from a microscopic treatment ~\cite{Chen2011,Levin2012} of their wavefunctions. This approach allowed us to re-derive various expected results, such as the edge theory and symmetry-flux response of type $\rn{2}$ SPTs. Concerning new results, it allowed us to derive a Laughlin-like picture for the $Z_{N>2}$ group-cohomology wavefunctions which includes quasi-long-range hidden order and a composite-particle interpretation. It further provided us with the entanglement spectrum of all SPTs we have considered. 

From the mathematical perspective, perhaps the most intriguing result is the direct connection between group cohomology cocycles and critical, sometimes integrable, models. Indeed, all ``Statistical Mechanics'' models we have obtained by putting products of group cohomology cocycles as the Boltzman weights, were critical and some were even integrable. Notably, even if one associated these models with the physical edges, criticality is by no means implied as generic edge theories of SPTs can be both critical or broken-symmetry phases. It is possible that this hidden criticality of the wave function amplitude is generic or that group cohomology wavefunctions are naturally tuned to be in the critical regime where CFT gives us a useful handle on their properties. Regardless, our results suggest an intriguing correspondence between group cohomology cocycles and critical lattice models. 

It is also noteworthy that the loop model related to one of the SPTs we looked at has a loop fugacity of $n=2$ and therefore breaks the ``$n=\sqrt{2}$ barrier'' which sets an upper bound on the loop fugacity of topological quantum loop models and has strongly hampered the realization of doubled $SU(2)_k$ topologically ordered states with loop models \cite{freedman2004class, Fendley20083113}.
This barrier arises from the fact that (1) loop models are critical only for $|n|\leq2$\cite{Jacobsen} and (2) the Stat. Mech. model considered within this construction is given by $Z=\sum_{\mathcal{C}} |\psi(\mathcal{C})|^2$, where $\mathcal{C}$ runs over loop configurations.
Crucially, as argued in this work, the relevant Stat. Mech. model for SPTs is given instead by $Z=\sum_{\mathcal{C}} \psi(\mathcal{C})$ and therefore only has a $n=2$ barrier.
It would be interesting to see how this discussion generalized to SPTs with $G=[Z_{N>2}]^3$, in which case loops are allowed to branch.

The current work suggests that a CFT approach to SPT phases can be unifying, useful for microscopics and that it can bring physical intuition about these states.
There are various directions along which it could be further explored.
It seems for example natural to generate fractional (i.e. long-range entangled) SPT phases by orbifolding the CFT with respect to a subgroup $G$. 
Such orbifolding would introduce magnetic operators with fractional charge into the CFT which would bind, via the symmetry-flux response argument, a fractional symmetry-charge. 
It would also be interesting to further explore the mapping between SPTs and integrable models introduced here, and even more so to study whether it may be reversed. 
This way, the sizable knowledge that was accumulated in studying integrable models could be used to obtain new microscopic models for SPT phases. 
Since the entanglement spectra we have obtained all had integer central charge, interesting candidates are integrable models with a fractional central charge. 
Furthermore, since the group cohomology approach becomes less comprehensive for fermionic SPTs \cite{PhysRevB.90.115141}, it may be useful to consider fermionic integrable models as a means of generating microscopic wavefunctions for fermionic SPTs.

\begin{acknowledgements}
Z. R. and T. S. would like to thank Steven Simon, Fabian Essler, John Cardy, Paul Fendley, Jesper Jacobsen, Romain Vasseur, Yuan-Ming Lu, Norbert Schuch and Curt von Keyserlingk for helpful discussions. T.S. was supported by the Clarendon Fund and Merton College and Z.R. was supported by an EPSRC grant and the European Union's Horizon 2020 research and innovation programme under the Marie Sklodowska-Curie grant agreement No. 657111. Both authors contributed equally to this work.
\end{acknowledgements}

\appendix

\section{Numerical evidence for the identification of a free boson CFT}
\label{Sec:NumericalBosons}
\subsection{Abelian}
Here we give numerical evidence which strongly supports the conjecture made in the text, that the compact boson CFT is the long distance theory governing the $Z_2$ and $Z_3$ group cohomology wavefunctions. 
To this end, we consider the statistical mechanics problem associated with the $A_{ \{ \phi_r \}}$ ``Boltzmann weights", given the hexagonal branching structure used in the main text and the $A$ sublattice analytical integration. 
The partition function ($Z$) then simply counts the number of configurations on an hexagonal lattice which obey the zero discrete vorticity condition (${\rm d}_2(r_a) = 0$). 

To analyze the CFT operator content of this problem, we use the transfer matrix approach \cite{Cardy1986}. We consider the hexagonal lattice as a brick lattice with the straight lines being vertical (i.e. the bricks are laid vertically). A transfer matrix ($T_l$) is then generated such that $\Tr[T_l^{L/2}]$ gives the partition function of a torus of circumference $l$ and length $L$, where the torus consists of $L$ vertical lines, each containing $l$ sites. The negative logarithm of the eigenvalues of $T_l$ ($\lambda(l)_i$), when normalized according to $\Lambda(l)_i = \lambda(l)_i \frac{L}{(2 \pi)(2\sqrt{3})}$ and shifted such that $\Lambda(l)_0 = 0$, then correspond directly, in the limit of large $l$, to the scaling dimension ($\Delta_i$) of different operators in the CFT. 

The primary operators of a compact boson CFT are given by the vertex operators $V_{e,m}$ with $e,m \in \mathbb{Z}$, where $e$ is the electric charge and $m$ the magnetic charge.
The operator content consists of these primary operators and of their descendants.
For the microscopic model studied here, we expect both electric and magnetic charges to be present since, as argued in the main text, charge operators on the $A$ sublattice generate magnetic charges and charges on the $B$ and $C$ sublattice generate electric charges. 
%
With a compactification radius of $1$, the scaling dimensions of primary operators as a function of $g$ are $\Delta_{e,m} = \frac{e^2}{2g} + \frac{g m^2}{2}$. 
Each of these vertex operators has descendants, or equivalently particle-hole excitations of the compact boson \cite{DiFran}. 
For our analysis we shall only require the two lowest lying of such excitations ($n=\pm 1$), with scaling dimension $\Delta_{e,m}  + 1$.
The 7 most relevant scaling dimensions $\Delta_i$, assuming $g$ smaller than but close to $1$, are given by $\Delta=0$ for the identity $I$, $\Delta = g/2$ for the two unit magnetic charges $V_{m=\pm 1}$, $\Delta=1/2g$ for the two unit electric charges $V_{e=\pm1}$, $\Delta=1$ for the two level one descendants of the identity $I \otimes (n=\pm 1)$, $\Delta=g/2+1/2g$ for the four combinations of $V_{e=\pm1,m=\pm1}$, $\Delta=g/2+1$ for the four level one descendants of the unit magnetic charge $V_{m=\pm1}\otimes (n = \pm 1)$ and lastly $\Delta=1/2g+1$ for the four level one descendants of the unit electric charge $V_{e=\pm1} \otimes (n \pm 1)$.
This sums up to a number of 19 operators for the 7 most relevant scaling dimensions.
Table~\ref{Table:Z2} lists the first 19 values of $\Lambda_i$, for the $Z_2$ group-cohomology wavefunction with the hexagonal branching structure, and $l=16,18,20$. The degeneracy of eigenvalues is given in the brackets, when it is different from $1$. 
This table agrees well with $g=1$, with discrepancies of the order of $10^{-3}$ which are attributed to finite size effects. 
Table~\ref{Table:Z3} lists the first 19 eigenvalues of $\Lambda_i$ for the $Z_3$ group-cohomology wavefunction with the hexagonal branching structure, and $l=10,12,14$. 
We find that the optimal fit for these eigenvalues is given by $g = 0.925 \pm 0.01$.

One can also extract the central charges of these two CFTs by fitting the lowest $\lambda(l)_i$ to $-s_0 \cdot l - \frac{\sqrt{3}}{2}(\pi c)/(6\cdot l)$ \cite{Cardy1986}. In the $Z_2$ case we find $c=0.9983(1)$, and in the $Z_3$ case we find $c=0.997(1)$.  

Lastly we comment on several 2-point and 4-point correlation functions that we obtained on a cylindrical geometry. For the $Z_3$ case we find that charge neutrality has to be obeyed on the $A$ sublattice and $B+C$ sublattices separately, such that for example, the two point function between an ${\rm O}_{+}$ operator on $A$ and an ${\rm O}_{-}$ operator on $B$ is always zero. This is expected due to the extra $Z_N$ symmetry implied by the hexagonal branching structure. Furthermore we find that $\langle {\rm O}_{+}(r) {\rm O}_{-}(r') \rangle$ is always positive when both are placed on the $B$ sublattice (or when both are placed on the $C$ sublattice) and always negative when one is on the $B$ sublattice and the other on the $C$ sublattice. Also considered are four point function $\langle {\rm O}_+(r_a) {\rm O}_{+}(r_b) {\rm O}_{-}(r'_a) {\rm O}_{-}(r'_b) \rangle$, on a cylindrical geometry, where by $r_{a/b/c}$ we denote both the position and the sublattice. Winding $r_b$ around $r_a$, we find that the complex phase of the correlator rotated by $2 \pi$ as expected. This behavior agrees very well with CFT predictions given the association between lattice operators and CFT operators made in the main text. 

For the $Z_2$ case, due to the extra $Z_2 \times Z_2 \times Z_2$ symmetry, we find that charge neutrality has to be obeyed on each sublattice separately. 
In agreement with this fact and the two assumptions given in the main text, the following correspondence between microscopic and CFT operators is proposed: on the $B$ ($C$) sublattice, $O \rightarrow V_{e=+1} + V_{e=-1}$ ($O \rightarrow i(V_{e=+1} - V_{e=-1})$) and on the $A$ sublattice, $O \rightarrow V_{m=+1} + V_{m=-1}$.
This correspondence is again consistent with the previously mentioned numerical results.


\begin{table*}[t]
    \centering 
    \begin{tabular}{ |c | c | c | c | c | c |}
    \hline
    CFT Operator & $\Delta(g=1)$ & $\Lambda(l=16)$ & $\Lambda(l=18)$ & $\Lambda(l=20)$ \\ \hline
    $I$ & 0 & 0 & 0 & 0 \\ \hline
    $V_{m = \pm 1}$ &0.5(x4) & 0.50235616(x4) & 0.50186276(x4) & 0.50150942(x4)  \\
    $V_{e = \pm 1}$ & & & & \\  \hline 
    $I\otimes(n = \pm 1)$ & 1(x6) & 0.99901361(x6) & 0.9989935 (x6) & 0.99905296(x6)   \\ 
    $V_{e=\pm1,m=\pm1}$ &  & & &  \\ \hline 
     $V_{m = \pm 1}\otimes(n = \pm 1)$ &1.5(x8) & 1.51518252(x8)  & 1.51182344(x8) & 1.50947823(x8)   \\ 
    $V_{e = \pm 1}\otimes(n = \pm 1)$ &  & & &    \\ 
    \hline
    \end{tabular} 
    \caption{   \label{Table:Z2}{\bf Comparison between numerically obtained scaling dimensions and those of a compact boson with $g=1$, for the $Z_2$ group cohomology wavefunction.}}
\end{table*}

\begin{table*}[t]
    \centering 
    \begin{tabular}{ |c | c | c | c | c | c |}
    \hline
    CFT Operator & $\Delta(g=0.925)$ & $\Lambda(l=10)$ & $\Lambda(l=12)$ & $\Lambda(l=14)$ \\ \hline
    $I$ & 0 & 0 & 0 & 0 \\ \hline
    $V_{m = \pm 1}$ & 0.4625(x2) & 0.46582886(x2) & 0.46457917(x2) & 0.46382596(x2)  \\ \hline
    $V_{e = \pm 1}$ & 0.540540(x2) & 0.54756391(x2) & 0.54568348(x2) & 0.5445524(x2)  \\ \hline
    $I\otimes(n = \pm 1)$ & 1(x2) & 1.0031008(x2) & 1.00042102(x2) & 0.99954883(x2)   \\ \hline
    $V_{e=\pm1,m=\pm1}$ & 1.00304054(x4) & 1.00195587(x4) & 1.00156733(x4) & 1.0016539(x4)   \\ \hline
    $V_{m = \pm 1}\otimes(n = \pm 1)$ & 1.4625(x4) & 1.49370415(x4) & 1.48280295(x4) & 1.47672672(x4)   \\ \hline
    $V_{e = \pm 1}\otimes(n = \pm 1)$ & 1.54054054(x4) & 1.58217044(x4) & 1.56804518(x4) & 1.56029936(x4)   \\ 
    \hline
    \end{tabular} 
     \caption{ \label{Table:Z3}{\bf Comparison between numerically obtained scaling dimensions and those of a compact boson with $g=0.925$, for the $Z_3$ group cohomology wavefunction.}}
\end{table*}

\subsection{Non-Abelian}
We have performed the same transfer matrix numerical calculation for the case of $D_3$. The spectrum given in Table \ref{Table:D3} gives strong evidence for a free boson at $g=1$.

\begin{table*}[ht!]
    \centering 
    \begin{tabular}{ |c | c | c | }
    \hline
    CFT Operator & $\Delta(g=1)$ & $\Lambda(l=6)$ \\ \hline
    $I$ & 0 & 0 \\ \hline
    $V_{m = \pm 1}$ &0.5(x4) & 0.518(0),  0.511(9),  0.513(0),  0.514(7)  \\
    $V_{e = \pm 1}$ & &  \\  \hline 
    $I\otimes(n = \pm 1)$ & 1(x6) & 0.949(7) (x2), 0.948(3) (x2),  0.949(0) (x2)  \\ 
    $V_{e=\pm1,m=\pm1}$ &  &  \\
    \hline
    \end{tabular} 
    \caption{   \label{Table:D3}{\bf Comparison between numerically obtained scaling dimensions and those of a compact boson with $g=1$, for the $D_3$ group cohomology wavefunction with a type 1 cocycle with $p=1$.}}
\end{table*}

\section{Operator content and association for the $c=-7$ loop model}
\label{Sec:NumericalLoops}
\begin{table*}[t]
    \centering 
    \begin{tabular}{ |c | c | c | c | c | c |}
    \hline
    CFT Operator & $\Delta(c=-7)$ & $\Lambda(l=18)$ & $\Lambda(l=21)$ & $\Lambda(l=24)$ \\ \hline
    $V_{e=e_0}$ & -2/3 & -2/3 & -2/3 & -2/3 \\ \hline
    $W_{l=1}$ & -1/2(x2) &-0.50090528(x2) & -0.50067042(x2) & -0.50051597(x2)  \\ 
    $V_{e=1}$ & & & & \\ \hline 
    $V_{e=0} \equiv I$ & 0(x2) & 0.00195229(x2) & 0.00141917(x2) & 0.00107904(x2)  \\ 
    $W_{l=2} $ &  & &  &    \\ \hline
    $(\partial_z,\partial_{\bar{z}}) \cdot V_{e=e_0}$ & 1/3(x2) & 0.35635165(x2) & 0.35013343(x2) & 0.34614146(x2)   \\ \hline
     \end{tabular} 
     \caption{\label{TableLG}{\bf Comparison between numerically obtained scaling dimensions for the Levin-Gu wavefunction and those of the Coulomb gas predictions for the $c=-7$ CFT.}}
\end{table*}

Here we analyze the operator content of the $n=-1$ dense loop model corresponding to the Levin-Gu wavefunction. Our aim is both to identify which CFT operators appear in this specific microscopic model as well as understand the association between this set of operators and the microscopic charge operators. To this end, we repeat the transfer matrix analysis used in the previous appendix, however this time on the full triangular lattice. The transfer matrix ($T$) generates the partition function ($Z$) for a cylinder of circumference $l$ with one of the two primitive vectors of the lattice oriented along the circumference, such that straight lines of sites encircle it. The transfer matrix ``propagates'' along two such horizontal lines such that it does not mix the three sublattices. To allow a consistent labelling between the sublattices, $l$ was taken to be a multiple of $3$.

The amplitude $A_{ \{\sigma^z_r \}}$ is simply given by $(-1)$ to the number of domain walls in the $\sigma^z_r$ configuration.
Since domain walls are not intersecting for $Z_2$ spins on a triangular lattice, they can be interpreted as loops and their number is well defined. 
The resulting 2D statistical mechanics problem is a special case ($n=-1$, $x=1$) of a general loop model whose partition function reads
\begin{equation}
Z = \sum_{\mathcal{C}} n^{\#[\mathcal{C}]} x^{l[\mathcal{C}]}
\end{equation}
where $\mathcal{C}$ is a loop configuration, $\#[\mathcal{C}]$ is the number of loops, $l[\mathcal{C}]$ is the total length of loops and $x$ is the inverse loop tension.
Under renormalization, it flows to a stable critical phase (the so-called dense-loop phase) with $x=x_0 \equiv (2-\sqrt{2-n})^{-1/2}$\cite{Nienhuis1982,Cardy2000,Jacobsen2003,Jacobsen,Cenke2014}. 

This critical phase has a central chage of $c=-7$ and is a non-unitary logarithmic CFT. Yet, based on the Coulomb gas approach, one expects primary operators to be some subset of electric charges ($V_e$) and $2l$-leg watermelon operators ($W_l$) whose scaling dimension in this case are $\Delta_{e} = e(e-2 e_0)/2g$ and $\Delta_{l} = g l^2/2 - e_0^2/2g$, with $e_0 = 2/3$ and $g=1/3$\cite{Nienhuis1982,Cardy2000,Jacobsen2003,Jacobsen}. 
In Table \ref{TableLG}, we find a good match with the numerically obtained scaling dimensions and the following subset of operators: $V_{e=e_0},W_{l=1},V_{e=1},V_{e=0} \equiv I,W_{l=2}$ and the level one descendants of $V_{e=e_0}$, denoted $(\partial_z,\partial_{\bar{z}}) \cdot V_{e=e_0}$. These calculations were performed for $x=x_0$.

Due to the presence of negative scaling dimensions, the previous method of extracting the central charge would now yield an effective central charge of $c_* = c - 12 \Delta_{\text{min}}$. Under the above identification of operators, $\Delta_{\text{min}} = -2/3$ and so we expect $c_* = -7+8 = 1$. Numerically we obtain $c_* = 0.98(9)$. 
It should be noted that, due to the presence of negative scaling dimensions, one has to be careful about the boundary conditions at the ends of the cylinder. Coulomb gas results are only expected if one takes the identity CFT operator $I$ as the boundary condition.

Calculating $\langle \sigma_z(0)\sigma_z(r)\rangle$ on long cylinders, we find an oscillation of the form $\cos(\frac{2 \pi}{3}r)$ ($r=0,1,2,...$) times an exponential decay (for $r>l$) with a decay length consistent with the scaling dimension of the operators $W_{l=1}$ and $V_{e=1}$.
Assuming $W_{l=1}$ and $V_{e=1}$ are related by the sublattice symmetry operations, the oscillation can be reproduced by assuming the following correspondence between microscopic and CFT operators: $O(r) \rightarrow \mathcal{O}_A \equiv V_{e=1}$ on the $A$ sublattice, $O(r) \rightarrow \mathcal{O}_B \equiv (-1/2) V_{e=1} + \sqrt{3}/2 W_{l=1}$ on the $B$ sublattice and $O(r) \rightarrow \mathcal{O}_C \equiv (-1/2) V_{e=1} - \sqrt{3}/2 W_{l=1}$ on the $C$ sublattice.

There is obviously some arbitrariness in this choice of correspondence.
This arbitrariness is related to the sublattice symmetries, as we now discuss, and has therefore no physical impact, as it should.
Let us define the $2\pi/3$ rotation $\mathcal{R}$.
At the microscopic level, one has $\mathcal{R}(O(r_A)) = O(r'_B)$ where $r'_B=\mathcal{R}(r_A)$ and where the index indicates the sublattice to whom $r$ belongs.
At the CFT level, the counterpart of this relation would be $\mathcal{R}(\mathcal{O}_A(r)) =  \mathcal{O}_B(r')$.
One can check from the correspondence above that this relation holds if one assumes that the two operators $V_{e=1}$ and $W_{l=1}$ form a real two-dimensional representation of the sublattice symmetry group such that $\mathcal{R}(V_{e=1}) = (-1/2) V_{e=1} + (\sqrt{3}/2) W_{l=1}$ and $\mathcal{R}(W_{l=1}) = (-\sqrt{3}/2) V_{e=1} - (1/2)  W_{l=1}$.
Further assuming that the sublattice symmetries commute with the conformal symmetry generators, one obtains that CFT correlators are invariant under the sublattice symmetries.





Unlike in the $Z_{N>2}$ group-cohomology case, the association between lattice and CFT operators in the Levin-Gu case is such that lattice operators appear as a linear combination of CFT primaries.
The expression for a particular amplitude written in the charge basis would thus appear as a superposition of many CFT correlators. 
This complication is however avoidable following a simple change of basis.
Pick one of the three main directions of the triangular lattice, say horizontal. For each horizontal $BC$ bond, replace the local basis states $\ket{B}$ and $\ket{C}$ by $|\pm\rangle = \frac{1}{\sqrt{2}} |B \rangle \pm |C\rangle$. Leave the basis on the $A$ sites as is.
Now, the previous correspondence implies that $O(r)$ for $|+\rangle$ is proportional to $V_{e}$ and $O(r)$ for $|-\rangle$ is proportional to $W_{l}$ while $O(r)$ for $|A\rangle$ is, as before, associated with $V_{e}$. 
Using this basis, a charge configuration now translates into a single CFT correlator.


    %

\section{The adiabatic connection with the dilute charge limit}
\label{Sec:NumericalDilute}
Here, evidence is given to show that the dilute charge limit, where non-zero charges are far apart on the lattice scale, is adiabatically connected to the dense charge limit. 
While the following approach is general, the numerical calculations were performed for the group cohomology~\cite{Chen2011} and the Levin-Gu~\cite{Levin2012} SPT wavefunctions.
Our approach goes as follows. 
First, we consider the parent Hamiltonians (${\rm H}_0$) of the original SPT wavefunctions\cite{Chen2011b,Chen2011,Levin2012}, which contains no particle fugacity factor yet and therefore correspond to the dense charge limit.
%
On closed surfaces, these Hamiltonians are local, gapped, and consist of a sum of commuting projectors (${\rm H}_0 = \sum_i {\rm h}_{0,i}$). 
In particular, they are frustration-free Hamiltonians \cite{Perez-Garcia2007}: each term in the sum is positive definite and annihilates the ground state. 
Consequently, ${\rm H}_0$ can also be thought of as the parent Hamiltonians \cite{Perez-Garcia2007} associated with the Tensor Product State (TPS) version of these wavefunctions $|\psi_0\rangle$, as shown in Eq.~\ref{Eq:SPTCFT}.

These TPS wavefunctions are then modified to $|\psi_{\beta}\rangle$ by introducing a fugacity of $\beta$ for any non-zero charge, such that $\beta=0$ forbids charges altogether and $\beta=1$ yields the original wavefunction ($\ket{\psi_{\beta=1}} = \ket{\psi_0}$). 
Formally, this is done by using a non-unitary transformation $\Lambda_{\beta}= \prod_r \beta^{\widehat{|\alpha_r|}}$, where $\widehat{|\alpha_r|}$ is the operator measuring the (minimal) absolute value of the charge at site $r$ (e.g. $\alpha_r = N-1$ leads to $\widehat{|\alpha_r|}=|-1|=1$) such that $|\psi_{\beta}\rangle = \Lambda_{\beta} |\psi_0\rangle$. For any finite $\beta$ this transformation has an inverse, and thus the following family of Hermitian local Hamiltonians can be defined: 
\begin{align}
{\rm H}_{\beta} &= \sum_i {\rm h}_{\beta,i} \\ \nonumber
{\rm h}_{\beta,i} &= \left(\lambda^{-1}_{\beta,i}\right)^{\dagger} {\rm h}_{*,i} \lambda^{-1}_{\beta,i},
\end{align}
where $\lambda_{\beta,i}$ is the product of $\beta^{|\alpha_r|}$ on the sites $r$ on which ${\rm h}_i$ acts non-trivially. 
It can be easily verified that ${\rm H}_{\beta}$ also consists of local positive-definite operators which annihilate $|\psi_{\beta}\rangle$ and thus is a parent Hamiltonian of $|\psi_{\beta}\rangle$. 
Except in the vicinity of $\beta=1$, the Hamiltonians ${\rm H}_{\beta}$ are not guaranteed to be gapped. 
Nonetheless, provided that their ground states $|\psi_{\beta} \rangle$ remain short-range correlated, it appears highly likely that they would be gapped.

\begin{figure}[h!]
\includegraphics[width=87mm, clip=true]{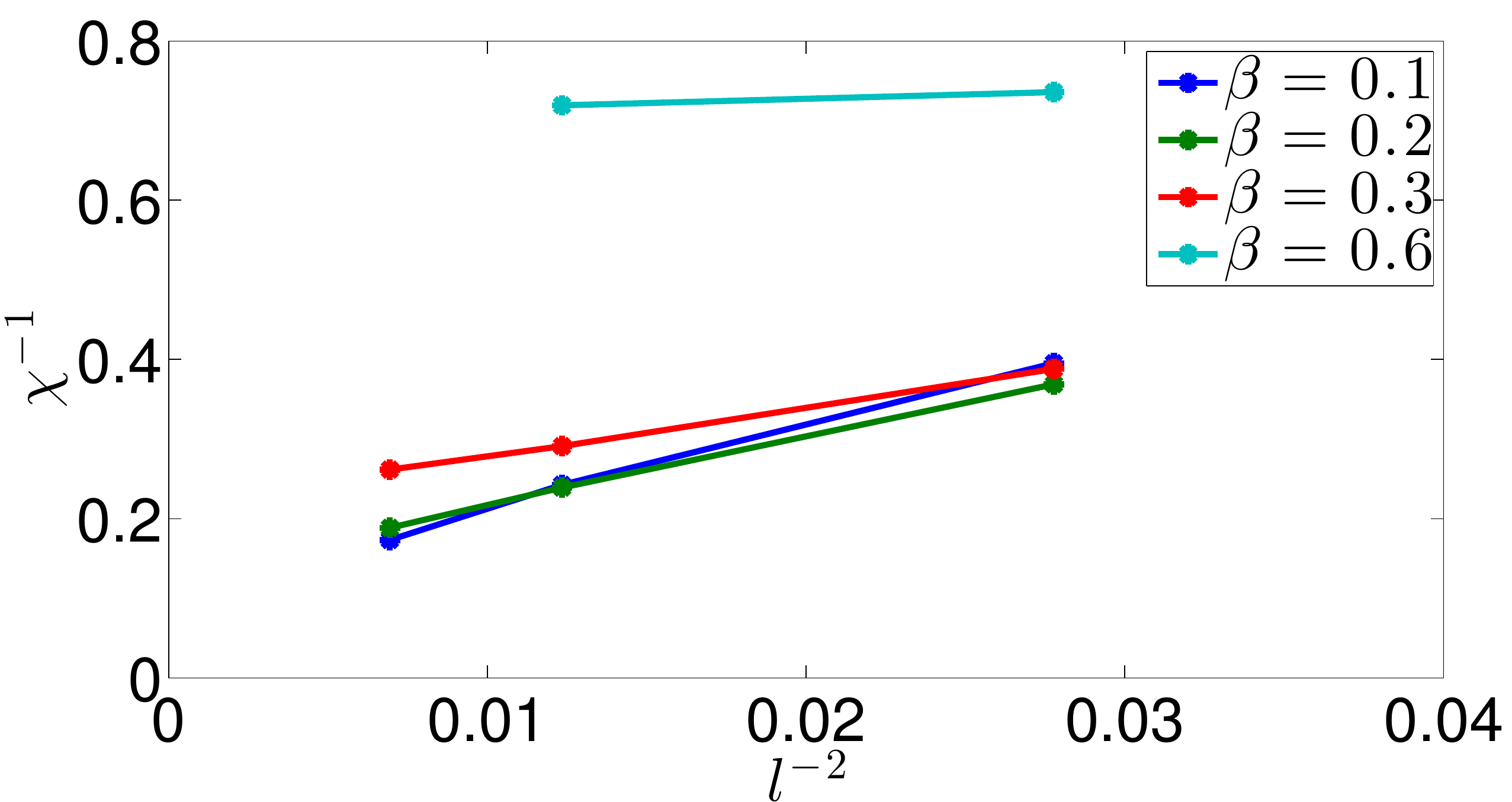}
\includegraphics[width=87mm, clip=true]{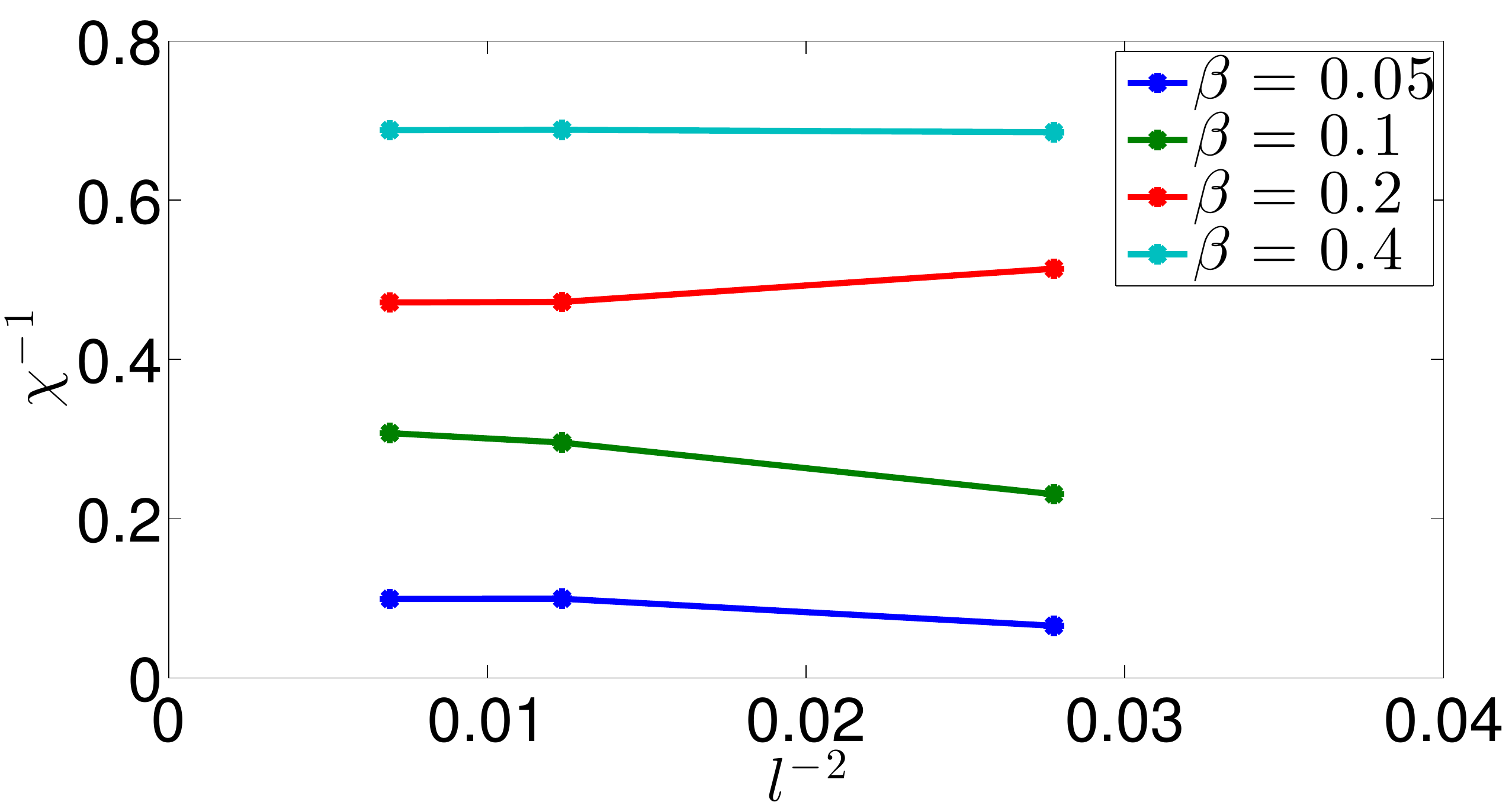}
\caption{Inverse correlation length ($\chi^{-1}$) for the $Z_2$ group-cohomology (upper panel) and Levin-Gu wavefunctions (lower panel) on a cylinder of circumference $l$ and for different values of $\beta$. 
For $Z_2$ group-cohomology, the correlation length appears to remain finite, at least down to $\beta=0.2$ where finite size correlations begin to interfere. For the Levin-Gu wavefunction, $\chi$ seems to decrease very rapidly with $\beta$, which makes it possible to do finite-size calculations for smaller values of $\beta$ than in the group-cohomology case.}
\label{Fig:Chi}
\end{figure}

The correlation length of these modified TPSs can be studied analytically at the two extremities of $\beta=0$ and $\beta=1$ and  by diagonalizing the TPS transfer operator (see Fig. (\ref{Fig:Chi})) at different circumferences ($l$) for general $\beta$.
Referring the reader to Ref.~\onlinecite{Verstraete2004} as regards the technicalities of this procedure, the resulting numerical computation for the $G=Z_2$ cases is easily described. 
One simply obtains two copies of the previously used partition function (${ Z},{ \bar{Z}}$), each with its own degrees of freedom ($\phi_r,\bar{\phi_r}$) and their corresponding ``Boltzmann weights'', ($A_{\{\phi_r\}}$,$A^*_{\{\bar{\phi}_r\}}$). 
On top of these phase factors, the resulting doubled sum over ``advanced" ($\phi_r$) and ``retarded" ($\bar{\phi}_r$) degrees of freedom now includes a factor of $(1+\beta \sigma_{r}^z \bar{\sigma}_{r}^z)$ for each site, where $\sigma_{r}^z = 1 - 2 \phi_r$. These terms represent the contraction of physical degrees of freedom between the retarded and advanced parts. Notably, at $\beta=1$ they force the retarded and advanced degrees of freedom to be locked together, while at $\beta=0$ they leave them decoupled. The correlation length of the wavefunction, with respect to any two local operators, is then bounded by the correlation length of this resulting ``bilayer'' statistical mechanics problem. 

For $\beta=0$ and $\beta=1$, this correlation length ($\chi$) can be analytically evaluated. Note that for $\beta=1$ the locking of retarded and advanced degrees of freedom removes all phases (since $A_{\{\phi_r\}} A^*_{\{\phi_r\}} = 1$), and one obtains a simple sum over only one set, say $\{\phi_r\}$, without any phases. 
This is a random spin partition function with $\chi=0$ as one expects from the original wavefunctions \cite{Chen2011}. 
In the limit $\beta \rightarrow 0$ the retarded and advanced parts become gradually decoupled and one obtains a doubled version of the CFTs studied in the main text. 
Consequently, $\chi$ tends to infinity. 
For small but finite $\beta$, a necessary (but not sufficient) condition to have a finite $\chi$ is that charge fugacity be a relevant perturbation for the doubled-CFT. 
Scaling dimensions in the doubled-CFT are just twice those of the previously studied ones, and so we find that the scaling dimension of $\sigma_z \otimes \bar{\sigma}_z$ is $\Delta_{\beta,LG}=-1$ and $\Delta_{\beta,Z2} = 1$ for the Levin-Gu and $Z_2$-group-cohomology wavefunctions, respectively. 
In both cases it is smaller than $2$ and consequently relevant as required. Based on this analysis, the simplest expectation for the phase diagram is a single phase transition happening exactly at $\beta=0$ with a finite correlation length for any $\beta>0$. Notably, the above argumentation generalizes to all $Z_N$-group-cohomology wavefunctions. 

To corroborate the above expectations, we numerically obtained $\chi^{-1}$ for cylinder circumferences of $l=6,9,12$ by taking the difference of the logarithms of the two dominant eigenvalue of the TPS transfer operator. 
The results for the original Levin-Gu wavefunction (i.e. with $x=1$) are shown in the lower panel of Fig. (\ref{Fig:Chi}) where the presence of a gap is evident even down to $\beta=0.05$. For the $Z_2$ group cohomology wavefunctions shown in the upper panel, $\chi$ appears to be generally larger thus limiting the regime were we can exclude a divergent $\chi$ to $\beta > 0.2$. The $x$ axis in Fig. (\ref{Fig:Chi}) was chosen to be one over the circumference squared ($1/l^2$) since it gives the best linear fit. If we had chosen to plot the gap as a function of $1/l$ instead, it would still appear that the gap remains finite above these values of $\beta$.

 \section{Corrections to the infinitely dilute charge limit}
 \label{Sec:DustyLaughlin}
The following expression was obtained for the $Z_{N>2}$ group cohomology wavefunction:
\begin{align}
A_{\{e_i,m_i,r_i\}} &= Z^{-1}\int D\varphi \prod_i V_{e_i,m_i}(r_i) e^{-S[\varphi]}.
\end{align}
The action is given by
\begin{equation}
\begin{aligned}
S[\varphi] &= \int d^2 r \frac{g}{4\pi} (\partial \varphi)^2 + \lambda (V_{e=N} + V_{e=-N}) \\
&+ \lambda' (V_{m=N} + V_{m=-N})
\end{aligned}
\end{equation}
where only the most relevant vertex terms allowed by symmetry were included.
In the main text, for simplicity, the infinitely dilute charge limit ($\rho \to 0$) was considered, in which case $\lambda,\lambda'\to 0$ since these vertex terms are irrelevant.
In that limit, since $A_{\{e_i,m_i,r_i\}}$ is non-zero only if $\sum_i e_i = \sum_i m_i =0$, the wavefunction possesses an enhanced $U(1)\times U(1)$ symmetry for the magnetic and electric charges instead of the microscopic $Z_N \times Z_N$ symmetry.

As we show below, at small but finite $\rho$, $\lambda,\lambda'$ become finite, and this acts to restore the original symmetry. Furthermore, we argue that the dominant configurations are the ones for which the excess charges ($\sum_i e_i = E,\sum_i m_i = M$) tend to bunch in tightly bound $N-$particle clouds whose core size is of the lattice scale. 
Under this picture, the wavefunction retains the factorizable form appearing in the main text and one can establish hidden order at all scales.

Let $l$ denote the average distance between charges, such that $\rho=a_0^2/l^2$. 
Between the lattice scale ($a_0$) and $l$, one may apply the usual RG procedure on the action, and find that $\lambda$ and $\lambda'$ are suppressed as $(a_0/l)^{-2+\Delta}$, with $\Delta$ being $N^2/2g$ and $g N^2/2$, respectively.
For neutral configurations ($E,M=0$), when charges are far apart on the lattice scale, the finiteness of $\lambda$ typically plays no role. 
The strong attraction between $+N$ and $-N$ charges would make them form neutral pairs on a microscopic scale \cite{KT} such that they would not interfere with the long range behavior. 

For a non-neutral configuration with $E=N$, $\lambda$ becomes essential for obeying charge neutrality in the CFT correlator, thus making the corresponding amplitude non-zero. 
This non-zero probability weight in the $E=kN$ sectors with $k\in\mathbb{Z}\setminus0$ restores the original $Z_N$ symmetry associated with electric particles.
In the following, we discuss an electric charge imbalance for simplicity, but the exact same argument applies to a magnetic charge imbalance.

For notational purposes, it is advantageous to enumerate the set $\{e_i,m_i,r_i\}$ by the positions $z^+_i$ of the $N_{+e}$ particles of charge $e=+1$, the positions $z^-_j$ of the $N_{-e}$ particles of charge $e=-1$, the positions $w^+_k$ of the $N_{+m}$ particles of charge $m=+1$ and  the positions $w^-_l$ of the $N_{-m}$ particles of charge $m=-1$.
With this notation, one has $A_{\{e_i,m_i,r_i\}} \equiv A(z^+_i;z^-_j;w^+_k;w^-_l)$, $E=N_{+e} -  N_{-e}$ and $M=N_{+m} -  N_{-m}$.
We now look at the wavefunction amplitudes in the $E=N$ sector.
At first order in $\lambda$, they are given by
\begin{equation}
\begin{aligned}
&A(z^+_i;z^-_j;w^+_k;w^-_l) = \int d^2 s \ \lambda \ \prod_{ijkl} \ \times\\ 
&\left\langle  V_{e=1}(z^+_i)V_{e=-1}(z^-_j)V_{m=1}(w^+_k)V_{m=-1}(w^-_l)V_{e=-N}(s)\right\rangle_0
\end{aligned}
\end{equation}
where $\left\langle \dots \right\rangle_0$ corresponds to taking the average for $\lambda,\lambda'=0$.
While the integrand can still be expressed in a Laughlin-like product form, the integration over the position of the charge $-N$ operator (henceforth referred to as the dust particle) makes the total amplitude unfactorizable. 

Notwithstanding, we conjecture that {\it typically} the above amplitude is in fact factorizable and retains a simple Laughlin-like form. As argued below, typical configurations are ones in which the dust particle is surrounded by a tight cloud made out of the excess $E=N$ charges. The core of this cloud is of the size of the microscopic lattice scale ($a_0$) and the integration over the position of the dust is effectively limited to this microscopic region and may be removed. Furthermore, the core of the cloud renormalizes the charge of the dust particle down to some effective value $N_{eff}$. As a result, the Laughlin-like form for the wavefunction is restored in a ``renormalized" form with new degrees of freedom consisting of the renormalized dust particle and all the original physical particles, excluding those in the cloud's core. Consequently, the arguments for hidden order used in the main text can be again applied using the plasma derived from this renormalized wavefunction. 

To support the above conjecture, we wish to show that it is self-consistent in terms of what the dominant configurations are.
Dominant configurations correspond to a large magnitude $|A(z^+_i;z^-_j;w^+_k;w^-_l)|$.
Let us assume this magnitude can be written as 
%
\begin{equation}
\begin{aligned}
\label{Eq:AtoPlasma}
&|A(z^+_i;z^-_j;w^+_k;w^-_l)| = \int d^2 s \ \lambda \   \prod_{ijkl} \ \times \ \\ 
&\left| \left\langle V_{e=1}(z^+_i)V_{e=-1}(z^-_j)V_{m=1}(w^+_k)V_{m=-1}(w^-_l)V_{e=-N}(s)\right\rangle\right| 
\end{aligned}
\end{equation}
which is allowed provided that the integral over $d^2s$ is dominated by a set of configurations for which the complex phase of $A(z^+_i;z^-_j;w^+_k;w^-_l)$ varies very little, as assumed previously.
We will then show that under this assumption, the dominant configurations of $|A(z^+_i;z^-_j;w^+_k;w^-_l)|$ are those in which it is indeed of this form. 

The main advantage of Eq.~\ref{Eq:AtoPlasma} is that the integrand in the right hand side now appears as the Boltzmann weight of two decoupled plasmas associated with the magnetic and electric particles. Relevant to our discussion is the electric plasma which, in plasma terminology\cite{Bonderson2011}, is a two-component plasma with $\Gamma=1/g$ and one dust particle of charge $-N$. For such values of $\Gamma<2$, the plasma is screening with a screening length of the order of $l$. Consequently, dominant configurations are those in which the excess charge form a screening cloud around the dust particle. 

Let us estimate the form of this screening cloud for large $N$. To start with, we consider the subsystem consisting of the (negative) dust particle and a single positive charge. The partition function of this subsystem is 
\begin{align}
Z_{\text{dust}+1} &= \int \frac{d^2 s}{a_0^2} \int \frac{d^2 z_1}{a_0^2} \frac{a_0^{N/g}}{|s-z_1|^{N/g}}.
\end{align} 
For $N/g > 2$, the ultraviolet divergence in the above partition function implies that the partition function is dominated by configurations where $z_1$ is bound to $s$ within a small microscopic region ($a_0$). 
Effectively, these particles are then paired into a composite object with a charge of $-N+1$. 
We may repeat the argument with another charge 1 particle which, provided the condition $(N-1)/g > 2$ is satisfied, will bind itself to the composite object.
Consequently, already at the microscopic scale, the dust charge will be strongly reduced down to $-N_{\text{eff}} = -\left \lfloor{2g}\right\rfloor$ since it would have $N-N_{\text{eff}}$ charge $+1$ particles bound to it.


We may now easily justify assumption \ref{Eq:AtoPlasma} self-consistently. 
A configuration $(z^+_i;z^-_j;w^+_k;w^-_l)$ is deemed typical if $N-N_{\text{eff}}$ of the $N_{+e}$ particles present at positions $z^+_i$ form a tightly bound cluster.
It is then advantageous to separate the set of charge $+1$ particles that form this cluster ($z^+_{i_p}$ with $p=1,\dots,N-N_{\text{eff}}$) from the other charge $+1$ particles ($z^+_{i_q}$ with $q=1,\dots,N_{+e}-N+N_{\text{eff}}$).
With this notation, one can write $|z^+_{i_p} - \tilde{z}| \simeq a_0 \ \forall \ p$ where $\tilde{z}$ is the center of mass of the positions $z^+_{i_p}$, i.e. $\tilde{z} = \sum_p z^+_{i_p} / (N-N_{\text{eff}})$.
For a typical configuration, the integral over $d^2 s$ is only appreciable within the bunching region, i.e. within a distance of the order of $a_0$ from $\tilde{z}$.
Since magnetic charges appear on the scale of $l\gg a_0$, typically they would not enter this small region. 
Consequently the variation of the complex phase of the integrand scales as $a_0/l$ within the effective region of integration. 
Thus, in the limit of small $a_0/l$, Eq.~\ref{Eq:AtoPlasma} is justified. 
Finally, the effective Laughlin-like form for $A(z^+_{i_p},z^+_{i_q};z^-_j;w^+_k;w^-_l)$ in the $E=N$ sector is simply given by 
\begin{equation}
\begin{aligned}
&A(z^+_{i_p},z^+_{i_q};z^-_j;w^+_k;w^-_l) = \psi(z^+_{i_p} - \tilde{z})  \prod_{i_qjkl} \times \\ 
&\langle V_{e=1}(z^+_{i_q})V_{e=-1}(z^-_j)V_{m=1}(w^+_k)V_{m=-1}(w^-_l)V_{e=-N_{eff}}(\tilde{z})\rangle  
\end{aligned}
\end{equation}
where $\psi(z^+_{i_p} - \tilde{z})$ is the short-scale wavefunction for the particles in the cluster and has no impact on the long-range behavior.
Now that the wavefunction amplitudes are factorized, it is possible to show hidden order in the same way as done in the main text.

\section{Relation to previous work}
As explained in the main introduction, some field theory arguments were given in order to identify the edge theories of SPT states for which a continuous space-time description applies.
Two different techniques were used.
The first one relies on embedding the symmetry group $G$ in (a product of several of) $U(1)$ group(s) and to study the associated Chern-Simons theories with unimodular $K$ matrices\cite{YuanMing2012,MengCheng2014}.
This technique is limited to 2+1D dimensions, Abelian groups and to cocycles that have a simple Chern-Simons interpretation (so-called type 1 and type 2 cocycles)\cite{MengCheng2014}.
For all the cases treated in these references, the obtained edge theory was a non-chiral free boson ($c=1$) with symmetry breaking terms. 
Since the relevance or irrelevance of these terms depends on the Luttinger parameter, which is non-universal and depends on the microscopics, this argument basically states the edge theory is either a free boson or a symmetry-broken phase.
Interestingly, our microscopic calculations show that, for the cases we have looked at, the Luttinger parameter is in a regime where the theory is critical, which was not guaranteed.

The second one relies on embedding the symmetry group $G$ in $SO(D+2)$ (for a SPT in $D$ spatial dimensions) and to look at a non-linear sigma model with $\theta=2\pi k$ with $k \in \mathbb{Z}$ \cite{PhysRevB.91.134404,Cenke2014}.
This technique is limited to groups that can be embedded in $SO(D+2)$.
There, for 2+1-dimensional SPTs, the predicted edge theory is $SU(2)_1$ plus some symmetry breaking terms.

We want to emphasize that, unlike these field theory techniques, (1) our technique applies deep in the SPT phase and (2) it applies to groups too big to be embedded in $SO(D+2)$ (like $Z_3^3$ for $D=2$), to non-Abelian groups and to type 3 cocycles.
We have at least one example ($G=Z_3^3$ with a type 3 cocycle) for which $c=2$ and for which it is not obvious how to deduce it from these field theory arguments.

\section{Branching structure}
Quite surprisingly, although edges of SPT states can be gapped (if they are symmetry-broken), all the entanglement spectra obtained here were gapless. Furthermore, as discussed earlier, some are even integrable. In general, one may wonder whether this is a universal property of such wave functions or, equivalently, whether the strange correlator of an SPT state with respect to a trivial state is always critical. While we still lack a comprehensive answer to this question, it seems that in all cases analyzed by us and others \cite{Cenke2014}, critical behavior was obtained. Notwithstanding, some changes to the wavefunction do result in changes to its corresponding CFT or at least to its convergence properties.

One example of such a change is a chance in the branching structure.
Instead of the $ABC$ branching structure used throughout this work, let us consider using a simpler branching structure for which the repeating unit is a parallelogram spanning four sites.
It can be shown that for the case of $G=Z_2$ and the choice of cocycle we have made, this has no effect. However for $G=Z_3$, it does alter the partition function. Repeating the numerics for such a branching structure still results in a critical theory (i.e. transfer matrix eigenvalues scale roughly as $1/L$) but the scaling dimensions seem far from converging to their asymptotic CFT values. For instance, descendants were not clearly distinguishable in the spectrum for the system sizes we could reach. 

If one were to alter the branching structure very sparsely, this behavior could be understood from the CFT perspective developed here. The $ABC$ branching structure resulted in an extra $Z_N$ symmetry acting only on the $A$ sublattice. This extra symmetry is removed by the altered branching structure. This allows terms such as $\cos(\phi-\theta)$ in the action having a scaling dimension of $1/2g+g/2$. 
These terms are relevant and should therefore drive the model away from the $c=1$ theory, to either a different critical theory or a gapped, symmetry-broken theory\cite{Tsvelik2001479}. Notably, for $Z_3$, we have verified that the altered branching structure still results in a critical theory. 

Thus the $ABC$ branching structure which we have used throughout this text, and probably also the cocycle itself, has an important consequence on both the analytical and numerical tractability of the resulting CFTs. While expected from the technical perspective, this is quite unusual from the mathematical perspective where this freedom of choice should have no physical consequences and in which there are no canonical choices. 

\section{Flux responses} 
\label{App:Flux}
In an Integer Quantum Hall phase with $\sigma_{xy}=e^2/h$, inserting half a flux quantum draws in half of an electron. Considering SPT phases based on type $\rn{2}$ cocycles, similar quantized responses to symmetry fluxes have been recently identified \cite{Juven2}. Here we show how this is reflected in our formalism. 

To study such effects, static symmetry fluxes should be introduced into the formalism. To this end, we consider a disk geometry, we add a (static) gauge field $A_l \in G$ on each link $l$ of lattice and we replace in the Hamiltonian all terms such as $\phi_i-\phi_j$ by $\phi_i-\phi_j+A_{l_{ij}}$, where $l_{ij}$ is the link connecting site $i$ to site $j$. 
Without fluxes, $A_{l}$ can be written as the discrete derivative of a scalar $f_i$ (i.e. $A_{l_{ij}} = f_j - f_i$) and can thus be removed using a local gauge transformation taking $\phi_i \rightarrow \phi_i - f_i$ on each site. A flux $g \in G$ at the centre of the disk means that $\sum_{l \in O} A_l = g$ where $O$ is any set of links surrounding the origin. Note also the trivial fact that every path that does not encircle the origin still obeys $\sum_{l \in O} A_l = 0$.

\begin{figure}[t]
\includegraphics[width=80mm, clip=true]{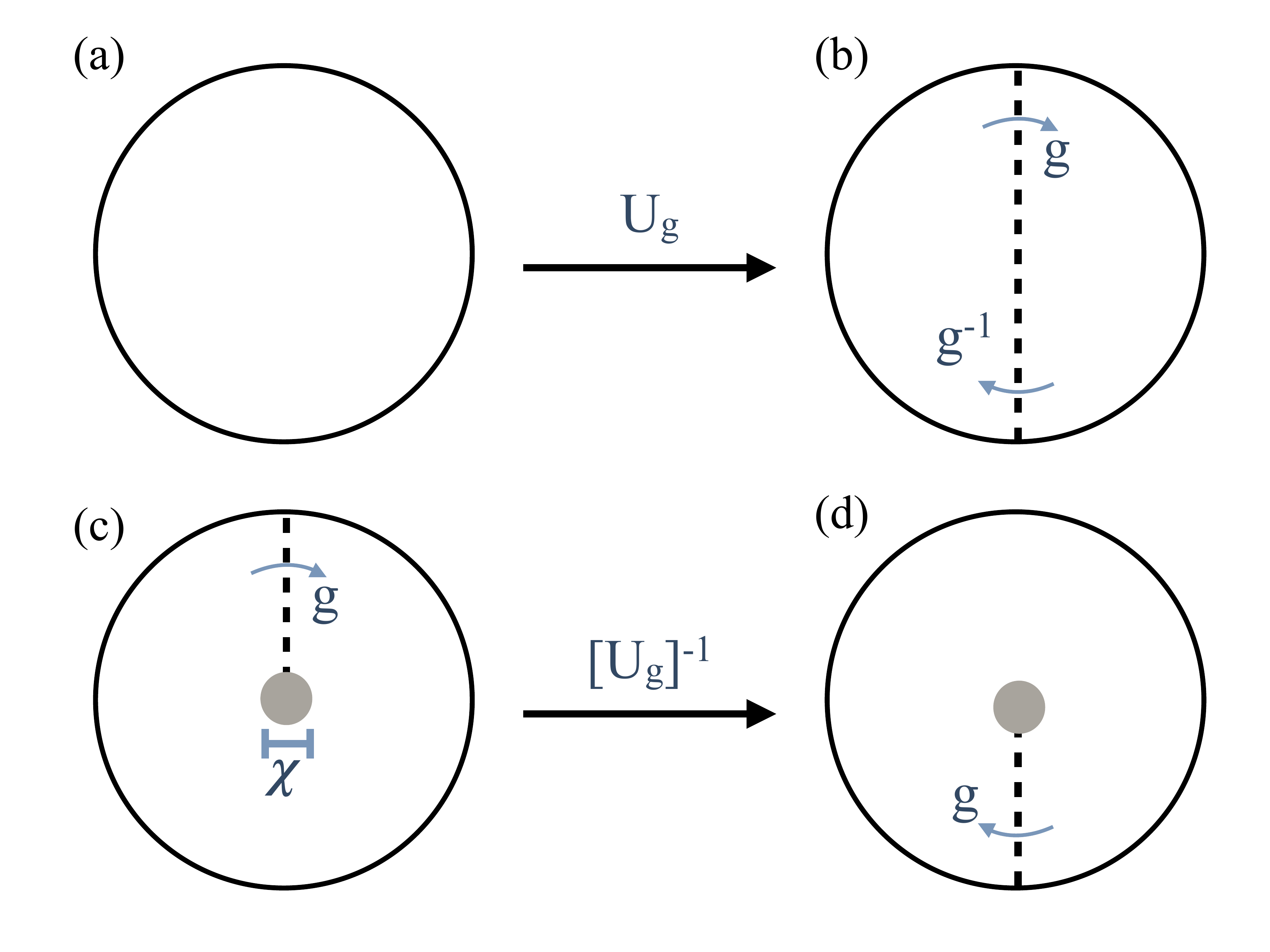}
\caption{Introducing static symmetry fluxes into the wave function. Acting with a unitary transformation which rotates by $g$ the right hand side of the system can be thought of as two opposite fluxes at the origin (see $(a)$ and $(b)$). Removing the lower flux yields the $g$-twisted wave function (see (c)). The location of the branch cut is insignificant and may be shifted using a unitary transformation (see (d)). }
\label{Fig:Flux}
\end{figure}

Next we wish to carry flux insertion from the Hamiltonian level to the ground state wavefunction level, or in other words relate the ground state wavefunctions before and after the flux insertion. Deferring proofs for a moment, the prescription for doing so is the following: First consider both $g$ and $g^{-1}$ fluxes, or equivalently act with a unitary transformation which rotates by $g$ all sites which are, say, on the right part of the disk (see Fig. (\ref{Fig:Flux})(a,b)). As shown in Section \ref{Sec:Type1} (see also Ref.~\onlinecite{YuanMing2012}), the physical symmetry simply rotates the fields of the CFT ($\varphi$ and its dual $\theta$) in a certain cocycle-dependent way. Thus the CFT obtained following this unitary transformation would simply have a $g$-twist in the boundary conditions of the CFT fields when crossing the upper vertical and $g^{-1}$-twist when crossing the lower line. Since the effects of the $g$-twist and of the $g^{-1}$-twist are distinct, one can now simply omit the $g^{-1}$-twist  (see Fig. (\ref{Fig:Flux})(c)). The resulting wave function is what we call the $g-$twisted groundstate wavefunction.

Next we explain how the $g$-twisted wave function is related to the groundstate of the Hamiltonian in the presence of a $g$ flux. Let us discuss the case which is of relevance here where the Hamiltonian is a sum of local projectors which annihilate the ground state. Introducing a $g$ flux in the Hamiltonian as in Fig. (\ref{Fig:Flux})(c) amounts to altering the projectors which act on both sides of this line via a Peierls-like substitution. Other projectors remain unaltered and thus a few correlation lengths ($\chi$) away from the $g-$flux insertion line, projectors acting on the $g$-twisted ground state still annihilate it. The reason is that, in a gapped system, all effects are local and thus the density matrix of the $g-$twisted groundstate in this region is exponentially close to the original ground state. 

Next we consider the subset of projectors which are close to the flux insertion line (on the scale of $\chi$) but still far from the origin. Here it is beneficial to use a unitary transformation and shift the flux line away from these projectors (see Fig (\ref{Fig:Flux})(d)). This removes the flux from both the projectors themselves and also, due to the finite correlation length, from the wave function. After this transformation, it is clear that the current subset of projectors again annihilates the $g$-twisted ground-state. However, since this transformation is unitary, it also means that, even prior to acting with it, the subset of projectors annihilated the $g-$twisted ground-state. Of course, it must then be that, also prior to this unitary transformation, it was annihilated by these projectors. We thus find that all projectors annihilate the $g-$twisted groundstate except perhaps those which are a few correlation lengths away from the flux insertion point. This implies that the $g$-twisted groundstate wave function captures the long range effect of the flux insertion (see also Ref. \onlinecite{Ringel2011}). 
 
Consider $G=Z_N \times Z_N$ with the simplest type $\rn{2}$ cocycle ($p_{\RN{1}}=p_{\RN{2}}=0,p' = 1$) and pick $g$ which acts only on the first group ($g=g_{\RN{1}} \times I$).  As shown in Section \ref{Sec:Type1}, within our CFT formulation the physical symmetry transformation $g$ rotates only the $\theta$ degree of freedom of the compact boson by $\frac{2\pi g}{N}$. The $g$-twisted wavefunction thus has a CFT with a $g-$twist only for the $\theta$ field. A few correlation lengths away from the flux insertion points, the effect of the flux thus appears as an insertion of $1/N$ of the basic magnetic charge.

In the dilute charge limit, the effect of inserting a fractional magnetic charge can be deduced from the plasma analogy used in the main text. From the two two-component plasmas associated with the norm of the wave function, the electric one would remain unchanged. However the magnetic one would contain an external point charge at the origin with a fraction of the fundamental charge. This external charge would get screened by the physical charges. When measuring the charge within a large region around the origin (or more accurately the difference in this quantity before and after the flux insertion), the external point charge would not be counted and only its screening cloud would be picked up, yielding a fractional result. 

Considering a generic SPT phase with $G=Z_N$ and a type $\rn{1}$ cocycle, a complication arises. As shown in Section \ref{Sec:Type1}, the physical symmetry now rotates both $\varphi$ and $\theta$. Within the plasma picture, this amounts to adding both half a magnetic and half an electric charge at the origin. Both electric and magnetic plasmas would then screen the external charge relevant to them and two screening clouds would be formed: One carrying half an electric charge and one carrying half a magnetic charge. Since the physical symmetry measures their sum, no fractional response associated with $G=Z_N$ would be detected. Indeed, in Ref. (\onlinecite{Juven2}) only type $\rn{2}$ cocycles were associated with fractional responses. Its worth mentioning that the specific type $\rn{1}$ wave functions considered in this work, which enjoy an enhanced $Z_N \times Z_N$ symmetry, would still see fractional charges associated with this enhanced symmetry. In particular, measuring the total charge of the $Z_N$ symmetry acting on the $A$ sublattice would pick up only the magnetic screening cloud and thus show a similar fractionalization effect as before. 

Based on the above result, we conjecture that a residual quantization effect remains even if the enhanced $Z_N \times Z_N$ symmetry is broken down to only the diagonal $Z_N$. Let $n_{1,i}$ be the operator which counts symmetry charges associated with only one of the $Z_N$'s on site $i$. Let $N_{1,l}$ be the sum of $n_{1,i}$ within a radius $l \gg\chi$ around the flux insertion point. As previously argued, the difference of $N_{1,l}$ before and after the flux insertion ($\Delta N_{1,l}$) would be fractional for the enhanced symmetry. Next consider a perturbation $V$ breaking $Z_N\times Z_N$ down to its diagonal part and present the wave function as a superposition of $N_{1,\infty}$ eigenvalues. Then consider the operator $N_{1,l} \,\ \mod \,\ 1$. Importantly, perturbation theory in $V$ only generates corrections to $N_{1,l}$ that are integer, and therefore $\Delta (N_{1,l} \,\ \mod \,\ 1)$ would still be fractional.

\bibliography{PRX_resub}

\begin{thebibliography}{73}
\expandafter\ifx\csname natexlab\endcsname\relax\def\natexlab#1{#1}\fi
\expandafter\ifx\csname bibnamefont\endcsname\relax
  \def\bibnamefont#1{#1}\fi
\expandafter\ifx\csname bibfnamefont\endcsname\relax
  \def\bibfnamefont#1{#1}\fi
\expandafter\ifx\csname citenamefont\endcsname\relax
  \def\citenamefont#1{#1}\fi
\expandafter\ifx\csname url\endcsname\relax
  \def\url#1{\texttt{#1}}\fi
\expandafter\ifx\csname urlprefix\endcsname\relax\def\urlprefix{URL }\fi
\providecommand{\bibinfo}[2]{#2}
\providecommand{\eprint}[2][]{\url{#2}}

\bibitem[{\citenamefont{Chen et~al.}(2011)\citenamefont{Chen, Liu, and
  Wen}}]{Chen2011b}
\bibinfo{author}{\bibfnamefont{X.}~\bibnamefont{Chen}},
  \bibinfo{author}{\bibfnamefont{Z.-X.} \bibnamefont{Liu}}, \bibnamefont{and}
  \bibinfo{author}{\bibfnamefont{X.-G.} \bibnamefont{Wen}},
  \bibinfo{journal}{Phys. Rev. B} \textbf{\bibinfo{volume}{84}},
  \bibinfo{pages}{235141} (\bibinfo{year}{2011}),
  \urlprefix\url{http://link.aps.org/doi/10.1103/PhysRevB.84.235141}.

\bibitem[{\citenamefont{Chen et~al.}(2013)\citenamefont{Chen, Gu, Liu, and
  Wen}}]{Chen2011}
\bibinfo{author}{\bibfnamefont{X.}~\bibnamefont{Chen}},
  \bibinfo{author}{\bibfnamefont{Z.-C.} \bibnamefont{Gu}},
  \bibinfo{author}{\bibfnamefont{Z.-X.} \bibnamefont{Liu}}, \bibnamefont{and}
  \bibinfo{author}{\bibfnamefont{X.-G.} \bibnamefont{Wen}},
  \bibinfo{journal}{Phys. Rev. B} \textbf{\bibinfo{volume}{87}},
  \bibinfo{pages}{155114} (\bibinfo{year}{2013}),
  \urlprefix\url{http://link.aps.org/doi/10.1103/PhysRevB.87.155114}.

\bibitem[{\citenamefont{Levin and Gu}(2012)}]{Levin2012}
\bibinfo{author}{\bibfnamefont{M.}~\bibnamefont{Levin}} \bibnamefont{and}
  \bibinfo{author}{\bibfnamefont{Z.-C.} \bibnamefont{Gu}},
  \bibinfo{journal}{Phys. Rev. B} \textbf{\bibinfo{volume}{86}},
  \bibinfo{pages}{115109} (\bibinfo{year}{2012}),
  \urlprefix\url{http://link.aps.org/doi/10.1103/PhysRevB.86.115109}.

\bibitem[{\citenamefont{Hasan and Kane}(2010)}]{Hasan2010}
\bibinfo{author}{\bibfnamefont{M.~Z.} \bibnamefont{Hasan}} \bibnamefont{and}
  \bibinfo{author}{\bibfnamefont{C.~L.} \bibnamefont{Kane}},
  \bibinfo{journal}{Reviews of Modern Physics} \textbf{\bibinfo{volume}{82}},
  \bibinfo{pages}{3045} (\bibinfo{year}{2010}).

\bibitem[{\citenamefont{Rasche et~al.}(2013)\citenamefont{Rasche, Isaeva, Ruck,
  Borisenko, Zabolotnyy, B{\"u}chner, Koepernik, Ortix, Richter, and van~den
  Brink}}]{Rasche2013}
\bibinfo{author}{\bibfnamefont{B.}~\bibnamefont{Rasche}},
  \bibinfo{author}{\bibfnamefont{A.}~\bibnamefont{Isaeva}},
  \bibinfo{author}{\bibfnamefont{M.}~\bibnamefont{Ruck}},
  \bibinfo{author}{\bibfnamefont{S.}~\bibnamefont{Borisenko}},
  \bibinfo{author}{\bibfnamefont{V.}~\bibnamefont{Zabolotnyy}},
  \bibinfo{author}{\bibfnamefont{B.}~\bibnamefont{B{\"u}chner}},
  \bibinfo{author}{\bibfnamefont{K.}~\bibnamefont{Koepernik}},
  \bibinfo{author}{\bibfnamefont{C.}~\bibnamefont{Ortix}},
  \bibinfo{author}{\bibfnamefont{M.}~\bibnamefont{Richter}}, \bibnamefont{and}
  \bibinfo{author}{\bibfnamefont{J.}~\bibnamefont{van~den Brink}},
  \bibinfo{journal}{Nat Mater} \textbf{\bibinfo{volume}{12}},
  \bibinfo{pages}{422} (\bibinfo{year}{2013}),
  \urlprefix\url{http://dx.doi.org/10.1038/nmat3570}.

\bibitem[{\citenamefont{Dziawa et~al.}(2012)\citenamefont{Dziawa, Kowalski,
  Dybko, Buczko, Szczerbakow, Szot, {\L}usakowska, Balasubramanian, Wojek,
  Berntsen et~al.}}]{story2012}
\bibinfo{author}{\bibfnamefont{P.}~\bibnamefont{Dziawa}},
  \bibinfo{author}{\bibfnamefont{B.~J.} \bibnamefont{Kowalski}},
  \bibinfo{author}{\bibfnamefont{K.}~\bibnamefont{Dybko}},
  \bibinfo{author}{\bibfnamefont{R.}~\bibnamefont{Buczko}},
  \bibinfo{author}{\bibfnamefont{A.}~\bibnamefont{Szczerbakow}},
  \bibinfo{author}{\bibfnamefont{M.}~\bibnamefont{Szot}},
  \bibinfo{author}{\bibfnamefont{E.}~\bibnamefont{{\L}usakowska}},
  \bibinfo{author}{\bibfnamefont{T.}~\bibnamefont{Balasubramanian}},
  \bibinfo{author}{\bibfnamefont{B.~M.} \bibnamefont{Wojek}},
  \bibinfo{author}{\bibfnamefont{M.~H.} \bibnamefont{Berntsen}},
  \bibnamefont{et~al.}, \bibinfo{journal}{Nat Mater}
  \textbf{\bibinfo{volume}{11}}, \bibinfo{pages}{1023} (\bibinfo{year}{2012}),
  \urlprefix\url{http://dx.doi.org/10.1038/nmat3449}.

\bibitem[{\citenamefont{Buyers et~al.}(1986)\citenamefont{Buyers, Morra,
  Armstrong, Hogan, Gerlach, and Hirakawa}}]{Buyers1986}
\bibinfo{author}{\bibfnamefont{W.~J.~L.} \bibnamefont{Buyers}},
  \bibinfo{author}{\bibfnamefont{R.~M.} \bibnamefont{Morra}},
  \bibinfo{author}{\bibfnamefont{R.~L.} \bibnamefont{Armstrong}},
  \bibinfo{author}{\bibfnamefont{M.~J.} \bibnamefont{Hogan}},
  \bibinfo{author}{\bibfnamefont{P.}~\bibnamefont{Gerlach}}, \bibnamefont{and}
  \bibinfo{author}{\bibfnamefont{K.}~\bibnamefont{Hirakawa}},
  \bibinfo{journal}{Phys. Rev. Lett.} \textbf{\bibinfo{volume}{56}},
  \bibinfo{pages}{371} (\bibinfo{year}{1986}),
  \urlprefix\url{http://link.aps.org/doi/10.1103/PhysRevLett.56.371}.

\bibitem[{\citenamefont{{Liu} et~al.}(2014)\citenamefont{{Liu}, {Gu}, and
  {Wen}}}]{Wen2014}
\bibinfo{author}{\bibfnamefont{Z.-X.} \bibnamefont{{Liu}}},
  \bibinfo{author}{\bibfnamefont{Z.-C.} \bibnamefont{{Gu}}}, \bibnamefont{and}
  \bibinfo{author}{\bibfnamefont{X.-G.} \bibnamefont{{Wen}}},
  \bibinfo{journal}{ArXiv e-prints}  (\bibinfo{year}{2014}),
  \eprint{1404.2818}.

\bibitem[{\citenamefont{{Koch-Janusz} et~al.}(2015)\citenamefont{{Koch-Janusz},
  {Khomskii}, and {Sela}}}]{Maciek2015}
\bibinfo{author}{\bibfnamefont{M.}~\bibnamefont{{Koch-Janusz}}},
  \bibinfo{author}{\bibfnamefont{D.~I.} \bibnamefont{{Khomskii}}},
  \bibnamefont{and} \bibinfo{author}{\bibfnamefont{E.}~\bibnamefont{{Sela}}},
  \bibinfo{journal}{ArXiv e-prints}  (\bibinfo{year}{2015}),
  \eprint{1501.03512}.

\bibitem[{\citenamefont{Wang et~al.}(2015{\natexlab{a}})\citenamefont{Wang,
  Nahum, and Senthil}}]{PhysRevB.91.195131}
\bibinfo{author}{\bibfnamefont{C.}~\bibnamefont{Wang}},
  \bibinfo{author}{\bibfnamefont{A.}~\bibnamefont{Nahum}}, \bibnamefont{and}
  \bibinfo{author}{\bibfnamefont{T.}~\bibnamefont{Senthil}},
  \bibinfo{journal}{Phys. Rev. B} \textbf{\bibinfo{volume}{91}},
  \bibinfo{pages}{195131} (\bibinfo{year}{2015}{\natexlab{a}}),
  \urlprefix\url{http://link.aps.org/doi/10.1103/PhysRevB.91.195131}.

\bibitem[{\citenamefont{Lu and Vishwanath}(2012)}]{YuanMing2012}
\bibinfo{author}{\bibfnamefont{Y.-M.} \bibnamefont{Lu}} \bibnamefont{and}
  \bibinfo{author}{\bibfnamefont{A.}~\bibnamefont{Vishwanath}},
  \bibinfo{journal}{Phys. Rev. B} \textbf{\bibinfo{volume}{86}},
  \bibinfo{pages}{125119} (\bibinfo{year}{2012}),
  \urlprefix\url{http://link.aps.org/doi/10.1103/PhysRevB.86.125119}.

\bibitem[{\citenamefont{{Kapustin}}(2014)}]{Kapustin2014}
\bibinfo{author}{\bibfnamefont{A.}~\bibnamefont{{Kapustin}}},
  \bibinfo{journal}{ArXiv e-prints}  (\bibinfo{year}{2014}),
  \eprint{1403.1467}.

\bibitem[{\citenamefont{Schuch et~al.}(2011)\citenamefont{Schuch,
  P\'erez-Garc\'~ia, and Cirac}}]{Schuch2011}
\bibinfo{author}{\bibfnamefont{N.}~\bibnamefont{Schuch}},
  \bibinfo{author}{\bibfnamefont{D.}~\bibnamefont{P\'erez-Garc\'~ia}},
  \bibnamefont{and} \bibinfo{author}{\bibfnamefont{I.}~\bibnamefont{Cirac}},
  \bibinfo{journal}{Phys. Rev. B} \textbf{\bibinfo{volume}{84}},
  \bibinfo{pages}{165139} (\bibinfo{year}{2011}),
  \urlprefix\url{http://link.aps.org/doi/10.1103/PhysRevB.84.165139}.

\bibitem[{\citenamefont{{Xu} and {Senthil}}(2013)}]{Xu2013}
\bibinfo{author}{\bibfnamefont{C.}~\bibnamefont{{Xu}}} \bibnamefont{and}
  \bibinfo{author}{\bibfnamefont{T.}~\bibnamefont{{Senthil}}},
  \bibinfo{journal}{\prb} \textbf{\bibinfo{volume}{87}}, \bibinfo{eid}{174412}
  (\bibinfo{year}{2013}), \eprint{1301.6172}.

\bibitem[{\citenamefont{Bi et~al.}(2015)\citenamefont{Bi, Rasmussen, Slagle,
  and Xu}}]{PhysRevB.91.134404}
\bibinfo{author}{\bibfnamefont{Z.}~\bibnamefont{Bi}},
  \bibinfo{author}{\bibfnamefont{A.}~\bibnamefont{Rasmussen}},
  \bibinfo{author}{\bibfnamefont{K.}~\bibnamefont{Slagle}}, \bibnamefont{and}
  \bibinfo{author}{\bibfnamefont{C.}~\bibnamefont{Xu}}, \bibinfo{journal}{Phys.
  Rev. B} \textbf{\bibinfo{volume}{91}}, \bibinfo{pages}{134404}
  (\bibinfo{year}{2015}),
  \urlprefix\url{http://link.aps.org/doi/10.1103/PhysRevB.91.134404}.

\bibitem[{\citenamefont{{Lu} and {Vishwanath}}(2013)}]{YuanMing2013}
\bibinfo{author}{\bibfnamefont{Y.-M.} \bibnamefont{{Lu}}} \bibnamefont{and}
  \bibinfo{author}{\bibfnamefont{A.}~\bibnamefont{{Vishwanath}}},
  \bibinfo{journal}{ArXiv e-prints}  (\bibinfo{year}{2013}),
  \eprint{1302.2634}.

\bibitem[{\citenamefont{Hung and Wen}(2013)}]{WenEnriched2013}
\bibinfo{author}{\bibfnamefont{L.-Y.} \bibnamefont{Hung}} \bibnamefont{and}
  \bibinfo{author}{\bibfnamefont{X.-G.} \bibnamefont{Wen}},
  \bibinfo{journal}{Phys. Rev. B} \textbf{\bibinfo{volume}{87}},
  \bibinfo{pages}{165107} (\bibinfo{year}{2013}),
  \urlprefix\url{http://link.aps.org/doi/10.1103/PhysRevB.87.165107}.

\bibitem[{\citenamefont{{Cheng} et~al.}(2015)\citenamefont{{Cheng}, {Bi},
  {You}, and {Gu}}}]{Gu2015}
\bibinfo{author}{\bibfnamefont{M.}~\bibnamefont{{Cheng}}},
  \bibinfo{author}{\bibfnamefont{Z.}~\bibnamefont{{Bi}}},
  \bibinfo{author}{\bibfnamefont{Y.-Z.} \bibnamefont{{You}}}, \bibnamefont{and}
  \bibinfo{author}{\bibfnamefont{Z.-C.} \bibnamefont{{Gu}}},
  \bibinfo{journal}{ArXiv e-prints}  (\bibinfo{year}{2015}),
  \eprint{1501.01313}.

\bibitem[{\citenamefont{Fubini and Lutken}(1991)}]{Fubini1991}
\bibinfo{author}{\bibfnamefont{S.}~\bibnamefont{Fubini}} \bibnamefont{and}
  \bibinfo{author}{\bibfnamefont{C.~A.} \bibnamefont{Lutken}},
  \bibinfo{journal}{Modern Physics Letters A} \textbf{\bibinfo{volume}{06}},
  \bibinfo{pages}{487} (\bibinfo{year}{1991}).

\bibitem[{\citenamefont{Moore and Read}(1991)}]{MooreRead1991}
\bibinfo{author}{\bibfnamefont{G.}~\bibnamefont{Moore}} \bibnamefont{and}
  \bibinfo{author}{\bibfnamefont{N.}~\bibnamefont{Read}},
  \bibinfo{journal}{Nuclear Physics B} \textbf{\bibinfo{volume}{360}},
  \bibinfo{pages}{362 } (\bibinfo{year}{1991}), ISSN \bibinfo{issn}{0550-3213},
  \urlprefix\url{http://www.sciencedirect.com/science/article/pii/055032139190407O}.

\bibitem[{\citenamefont{Blok and Wen}(1992)}]{Blok1992615}
\bibinfo{author}{\bibfnamefont{B.}~\bibnamefont{Blok}} \bibnamefont{and}
  \bibinfo{author}{\bibfnamefont{X.}~\bibnamefont{Wen}},
  \bibinfo{journal}{Nuclear Physics B} \textbf{\bibinfo{volume}{374}},
  \bibinfo{pages}{615 } (\bibinfo{year}{1992}), ISSN \bibinfo{issn}{0550-3213},
  \urlprefix\url{http://www.sciencedirect.com/science/article/pii/055032139290402W}.

\bibitem[{\citenamefont{Nayak et~al.}(2008)\citenamefont{Nayak, Simon, Stern,
  Freedman, and Das~Sarma}}]{NayakRevMod2008}
\bibinfo{author}{\bibfnamefont{C.}~\bibnamefont{Nayak}},
  \bibinfo{author}{\bibfnamefont{S.~H.} \bibnamefont{Simon}},
  \bibinfo{author}{\bibfnamefont{A.}~\bibnamefont{Stern}},
  \bibinfo{author}{\bibfnamefont{M.}~\bibnamefont{Freedman}}, \bibnamefont{and}
  \bibinfo{author}{\bibfnamefont{S.}~\bibnamefont{Das~Sarma}},
  \bibinfo{journal}{Rev. Mod. Phys.} \textbf{\bibinfo{volume}{80}},
  \bibinfo{pages}{1083} (\bibinfo{year}{2008}),
  \urlprefix\url{http://link.aps.org/doi/10.1103/RevModPhys.80.1083}.

\bibitem[{\citenamefont{Zaletel and Mong}(2012)}]{Zaletel2012}
\bibinfo{author}{\bibfnamefont{M.~P.} \bibnamefont{Zaletel}} \bibnamefont{and}
  \bibinfo{author}{\bibfnamefont{R.~S.~K.} \bibnamefont{Mong}},
  \bibinfo{journal}{Phys. Rev. B} \textbf{\bibinfo{volume}{86}},
  \bibinfo{pages}{245305} (\bibinfo{year}{2012}),
  \urlprefix\url{http://link.aps.org/doi/10.1103/PhysRevB.86.245305}.

\bibitem[{\citenamefont{Hsieh et~al.}(2014)\citenamefont{Hsieh, Sule, Cho, Ryu,
  and Leigh}}]{Ryu2014}
\bibinfo{author}{\bibfnamefont{C.-T.} \bibnamefont{Hsieh}},
  \bibinfo{author}{\bibfnamefont{O.~M.} \bibnamefont{Sule}},
  \bibinfo{author}{\bibfnamefont{G.~Y.} \bibnamefont{Cho}},
  \bibinfo{author}{\bibfnamefont{S.}~\bibnamefont{Ryu}}, \bibnamefont{and}
  \bibinfo{author}{\bibfnamefont{R.~G.} \bibnamefont{Leigh}},
  \bibinfo{journal}{Phys. Rev. B} \textbf{\bibinfo{volume}{90}},
  \bibinfo{pages}{165134} (\bibinfo{year}{2014}),
  \urlprefix\url{http://link.aps.org/doi/10.1103/PhysRevB.90.165134}.

\bibitem[{\citenamefont{{Ringel} and {Simon}}(2014)}]{Ringel2015}
\bibinfo{author}{\bibfnamefont{Z.}~\bibnamefont{{Ringel}}} \bibnamefont{and}
  \bibinfo{author}{\bibfnamefont{S.~H.} \bibnamefont{{Simon}}},
  \bibinfo{journal}{ArXiv e-prints}  (\bibinfo{year}{2014}),
  \eprint{1410.0318}.

\bibitem[{\citenamefont{Sule et~al.}(2015)\citenamefont{Sule, Changlani,
  Maruyama, and Ryu}}]{Ryu2015}
\bibinfo{author}{\bibfnamefont{O.~M.} \bibnamefont{Sule}},
  \bibinfo{author}{\bibfnamefont{H.~J.} \bibnamefont{Changlani}},
  \bibinfo{author}{\bibfnamefont{I.}~\bibnamefont{Maruyama}}, \bibnamefont{and}
  \bibinfo{author}{\bibfnamefont{S.}~\bibnamefont{Ryu}},
  \bibinfo{journal}{Phys. Rev. B} \textbf{\bibinfo{volume}{92}},
  \bibinfo{pages}{075128} (\bibinfo{year}{2015}),
  \urlprefix\url{http://link.aps.org/doi/10.1103/PhysRevB.92.075128}.

\bibitem[{\citenamefont{Duivenvoorden and Quella}(2013)}]{Quella2013}
\bibinfo{author}{\bibfnamefont{K.}~\bibnamefont{Duivenvoorden}}
  \bibnamefont{and} \bibinfo{author}{\bibfnamefont{T.}~\bibnamefont{Quella}},
  \bibinfo{journal}{Phys. Rev. B} \textbf{\bibinfo{volume}{88}},
  \bibinfo{pages}{125115} (\bibinfo{year}{2013}),
  \urlprefix\url{http://link.aps.org/doi/10.1103/PhysRevB.88.125115}.

\bibitem[{\citenamefont{Else et~al.}(2013)\citenamefont{Else, Bartlett, and
  Doherty}}]{Else2013}
\bibinfo{author}{\bibfnamefont{D.~V.} \bibnamefont{Else}},
  \bibinfo{author}{\bibfnamefont{S.~D.} \bibnamefont{Bartlett}},
  \bibnamefont{and} \bibinfo{author}{\bibfnamefont{A.~C.}
  \bibnamefont{Doherty}}, \bibinfo{journal}{Phys. Rev. B}
  \textbf{\bibinfo{volume}{88}}, \bibinfo{pages}{085114}
  (\bibinfo{year}{2013}),
  \urlprefix\url{http://link.aps.org/doi/10.1103/PhysRevB.88.085114}.

\bibitem[{\citenamefont{Levin and Wen}(2005)}]{LevinWen2005}
\bibinfo{author}{\bibfnamefont{M.~A.} \bibnamefont{Levin}} \bibnamefont{and}
  \bibinfo{author}{\bibfnamefont{X.-G.} \bibnamefont{Wen}},
  \bibinfo{journal}{Phys. Rev. B} \textbf{\bibinfo{volume}{71}},
  \bibinfo{pages}{045110} (\bibinfo{year}{2005}),
  \urlprefix\url{http://link.aps.org/doi/10.1103/PhysRevB.71.045110}.

\bibitem[{\citenamefont{You et~al.}(2014)\citenamefont{You, Bi, Rasmussen,
  Slagle, and Xu}}]{Cenke2014}
\bibinfo{author}{\bibfnamefont{Y.-Z.} \bibnamefont{You}},
  \bibinfo{author}{\bibfnamefont{Z.}~\bibnamefont{Bi}},
  \bibinfo{author}{\bibfnamefont{A.}~\bibnamefont{Rasmussen}},
  \bibinfo{author}{\bibfnamefont{K.}~\bibnamefont{Slagle}}, \bibnamefont{and}
  \bibinfo{author}{\bibfnamefont{C.}~\bibnamefont{Xu}}, \bibinfo{journal}{Phys.
  Rev. Lett.} \textbf{\bibinfo{volume}{112}}, \bibinfo{pages}{247202}
  (\bibinfo{year}{2014}),
  \urlprefix\url{http://link.aps.org/doi/10.1103/PhysRevLett.112.247202}.

\bibitem[{\citenamefont{S\o{}rensen et~al.}(2005)\citenamefont{S\o{}rensen,
  Demler, and Lukin}}]{Sorensen2005}
\bibinfo{author}{\bibfnamefont{A.~S.} \bibnamefont{S\o{}rensen}},
  \bibinfo{author}{\bibfnamefont{E.}~\bibnamefont{Demler}}, \bibnamefont{and}
  \bibinfo{author}{\bibfnamefont{M.~D.} \bibnamefont{Lukin}},
  \bibinfo{journal}{Phys. Rev. Lett.} \textbf{\bibinfo{volume}{94}},
  \bibinfo{pages}{086803} (\bibinfo{year}{2005}),
  \urlprefix\url{http://link.aps.org/doi/10.1103/PhysRevLett.94.086803}.

\bibitem[{\citenamefont{Hafezi et~al.}(2007)\citenamefont{Hafezi, S\o{}rensen,
  Demler, and Lukin}}]{Hafezi2007}
\bibinfo{author}{\bibfnamefont{M.}~\bibnamefont{Hafezi}},
  \bibinfo{author}{\bibfnamefont{A.~S.} \bibnamefont{S\o{}rensen}},
  \bibinfo{author}{\bibfnamefont{E.}~\bibnamefont{Demler}}, \bibnamefont{and}
  \bibinfo{author}{\bibfnamefont{M.~D.} \bibnamefont{Lukin}},
  \bibinfo{journal}{Phys. Rev. A} \textbf{\bibinfo{volume}{76}},
  \bibinfo{pages}{023613} (\bibinfo{year}{2007}),
  \urlprefix\url{http://link.aps.org/doi/10.1103/PhysRevA.76.023613}.

\bibitem[{\citenamefont{Scaffidi and M\"oller}(2012)}]{Scaffidi2012}
\bibinfo{author}{\bibfnamefont{T.}~\bibnamefont{Scaffidi}} \bibnamefont{and}
  \bibinfo{author}{\bibfnamefont{G.}~\bibnamefont{M\"oller}},
  \bibinfo{journal}{Phys. Rev. Lett.} \textbf{\bibinfo{volume}{109}},
  \bibinfo{pages}{246805} (\bibinfo{year}{2012}),
  \urlprefix\url{http://link.aps.org/doi/10.1103/PhysRevLett.109.246805}.

\bibitem[{\citenamefont{Liu and Bergholtz}(2013)}]{Liu2013}
\bibinfo{author}{\bibfnamefont{Z.}~\bibnamefont{Liu}} \bibnamefont{and}
  \bibinfo{author}{\bibfnamefont{E.~J.} \bibnamefont{Bergholtz}},
  \bibinfo{journal}{Phys. Rev. B} \textbf{\bibinfo{volume}{87}},
  \bibinfo{pages}{035306} (\bibinfo{year}{2013}),
  \urlprefix\url{http://link.aps.org/doi/10.1103/PhysRevB.87.035306}.

\bibitem[{\citenamefont{Kapit and Mueller}(2010)}]{Kapit2010}
\bibinfo{author}{\bibfnamefont{E.}~\bibnamefont{Kapit}} \bibnamefont{and}
  \bibinfo{author}{\bibfnamefont{E.}~\bibnamefont{Mueller}},
  \bibinfo{journal}{Phys. Rev. Lett.} \textbf{\bibinfo{volume}{105}},
  \bibinfo{pages}{215303} (\bibinfo{year}{2010}),
  \urlprefix\url{http://link.aps.org/doi/10.1103/PhysRevLett.105.215303}.

\bibitem[{\citenamefont{Qi}(2011)}]{Qi2011}
\bibinfo{author}{\bibfnamefont{X.-L.} \bibnamefont{Qi}},
  \bibinfo{journal}{Phys. Rev. Lett.} \textbf{\bibinfo{volume}{107}},
  \bibinfo{pages}{126803} (\bibinfo{year}{2011}),
  \urlprefix\url{http://link.aps.org/doi/10.1103/PhysRevLett.107.126803}.

\bibitem[{\citenamefont{Wu et~al.}(2013)\citenamefont{Wu, Regnault, and
  Bernevig}}]{Wu2013}
\bibinfo{author}{\bibfnamefont{Y.-L.} \bibnamefont{Wu}},
  \bibinfo{author}{\bibfnamefont{N.}~\bibnamefont{Regnault}}, \bibnamefont{and}
  \bibinfo{author}{\bibfnamefont{B.~A.} \bibnamefont{Bernevig}},
  \bibinfo{journal}{Phys. Rev. Lett.} \textbf{\bibinfo{volume}{110}},
  \bibinfo{pages}{106802} (\bibinfo{year}{2013}),
  \urlprefix\url{http://link.aps.org/doi/10.1103/PhysRevLett.110.106802}.

\bibitem[{\citenamefont{Harper et~al.}(2014)\citenamefont{Harper, Simon, and
  Roy}}]{Harper2014}
\bibinfo{author}{\bibfnamefont{F.}~\bibnamefont{Harper}},
  \bibinfo{author}{\bibfnamefont{S.~H.} \bibnamefont{Simon}}, \bibnamefont{and}
  \bibinfo{author}{\bibfnamefont{R.}~\bibnamefont{Roy}},
  \bibinfo{journal}{Phys. Rev. B} \textbf{\bibinfo{volume}{90}},
  \bibinfo{pages}{075104} (\bibinfo{year}{2014}),
  \urlprefix\url{http://link.aps.org/doi/10.1103/PhysRevB.90.075104}.

\bibitem[{\citenamefont{Scaffidi and Simon}(2014)}]{Scaffidi2014}
\bibinfo{author}{\bibfnamefont{T.}~\bibnamefont{Scaffidi}} \bibnamefont{and}
  \bibinfo{author}{\bibfnamefont{S.~H.} \bibnamefont{Simon}},
  \bibinfo{journal}{Phys. Rev. B} \textbf{\bibinfo{volume}{90}},
  \bibinfo{pages}{115132} (\bibinfo{year}{2014}),
  \urlprefix\url{http://link.aps.org/doi/10.1103/PhysRevB.90.115132}.

\bibitem[{\citenamefont{Gurarie et~al.}(1997)\citenamefont{Gurarie, Flohr, and
  Nayak}}]{Gurarie:1997dw}
\bibinfo{author}{\bibfnamefont{V.}~\bibnamefont{Gurarie}},
  \bibinfo{author}{\bibfnamefont{M.}~\bibnamefont{Flohr}}, \bibnamefont{and}
  \bibinfo{author}{\bibfnamefont{C.}~\bibnamefont{Nayak}},
  \bibinfo{journal}{Nucl. Phys.} \textbf{\bibinfo{volume}{B498}},
  \bibinfo{pages}{513} (\bibinfo{year}{1997}), \eprint{cond-mat/9701212}.

\bibitem[{\citenamefont{{Willerton}}(2005)}]{Gerbs}
\bibinfo{author}{\bibfnamefont{S.}~\bibnamefont{{Willerton}}},
  \bibinfo{journal}{ArXiv Mathematics e-prints}  (\bibinfo{year}{2005}),
  \eprint{math/0503266}.

\bibitem[{\citenamefont{Henley}(2005)}]{Henley2005}
\bibinfo{author}{\bibfnamefont{C.~L.} \bibnamefont{Henley}},
  \bibinfo{journal}{Phys. Rev. B} \textbf{\bibinfo{volume}{71}},
  \bibinfo{pages}{014424} (\bibinfo{year}{2005}),
  \urlprefix\url{http://link.aps.org/doi/10.1103/PhysRevB.71.014424}.

\bibitem[{\citenamefont{Francesco et~al.}(1997)\citenamefont{Francesco,
  Mathieu, and Senechal}}]{DiFran}
\bibinfo{author}{\bibfnamefont{P.~D.} \bibnamefont{Francesco}},
  \bibinfo{author}{\bibfnamefont{P.}~\bibnamefont{Mathieu}}, \bibnamefont{and}
  \bibinfo{author}{\bibfnamefont{D.}~\bibnamefont{Senechal}},
  \emph{\bibinfo{title}{Conformal Field Theory}}, Graduate Texts in
  Contemporary Physics (\bibinfo{publisher}{Springer}, \bibinfo{year}{1997}),
  ISBN \bibinfo{isbn}{9780387947853},
  \urlprefix\url{http://books.google.co.uk/books?id=keUrdME5rhIC}.

\bibitem[{\citenamefont{{Ginsparg}}(1991)}]{Ginsparg}
\bibinfo{author}{\bibfnamefont{P.}~\bibnamefont{{Ginsparg}}},
  \bibinfo{journal}{ArXiv High Energy Physics - Theory e-prints}
  (\bibinfo{year}{1991}), \eprint{hep-th/9108028}.

\bibitem[{\citenamefont{Senthil and Levin}(2013)}]{Senthil2013}
\bibinfo{author}{\bibfnamefont{T.}~\bibnamefont{Senthil}} \bibnamefont{and}
  \bibinfo{author}{\bibfnamefont{M.}~\bibnamefont{Levin}},
  \bibinfo{journal}{Phys. Rev. Lett.} \textbf{\bibinfo{volume}{110}},
  \bibinfo{pages}{046801} (\bibinfo{year}{2013}),
  \urlprefix\url{http://link.aps.org/doi/10.1103/PhysRevLett.110.046801}.

\bibitem[{\citenamefont{{de Wild Propitius}}(1995)}]{deWild}
\bibinfo{author}{\bibfnamefont{M.}~\bibnamefont{{de Wild Propitius}}}, Ph.D.
  thesis, \bibinfo{school}{PhD Thesis, 1995} (\bibinfo{year}{1995}).

\bibitem[{\citenamefont{Chen et~al.}(2014)\citenamefont{Chen, Lu, and
  Vishwanath}}]{AshvinDecorated}
\bibinfo{author}{\bibfnamefont{X.}~\bibnamefont{Chen}},
  \bibinfo{author}{\bibfnamefont{Y.-M.} \bibnamefont{Lu}}, \bibnamefont{and}
  \bibinfo{author}{\bibfnamefont{A.}~\bibnamefont{Vishwanath}},
  \bibinfo{journal}{Nat Commun} \textbf{\bibinfo{volume}{5}}
  (\bibinfo{year}{2014}), \urlprefix\url{http://dx.doi.org/10.1038/ncomms4507}.

\bibitem[{\citenamefont{Jacobsen}(2009)}]{Jacobsen}
\bibinfo{author}{\bibfnamefont{J.}~\bibnamefont{Jacobsen}},
  \emph{\bibinfo{title}{Polygons, Polyominoes and Polycubes}}
  (\bibinfo{publisher}{Springer Netherlands}, \bibinfo{year}{2009}), chap.
  \bibinfo{chapter}{Conformal Field Theory Applied to Loop Models}, Lecture
  Notes in Physics, ISBN \bibinfo{isbn}{9781402099274}.

\bibitem[{\citenamefont{Coste et~al.}(2000)\citenamefont{Coste, Gannon, and
  Ruelle}}]{ModularData}
\bibinfo{author}{\bibfnamefont{A.}~\bibnamefont{Coste}},
  \bibinfo{author}{\bibfnamefont{T.}~\bibnamefont{Gannon}}, \bibnamefont{and}
  \bibinfo{author}{\bibfnamefont{P.}~\bibnamefont{Ruelle}},
  \bibinfo{journal}{Nuclear Physics B} \textbf{\bibinfo{volume}{581}},
  \bibinfo{pages}{679 } (\bibinfo{year}{2000}), ISSN \bibinfo{issn}{0550-3213},
  \urlprefix\url{http://www.sciencedirect.com/science/article/pii/S0550321300002856}.

\bibitem[{\citenamefont{Nienhuis}(1982)}]{Nienhuis1982}
\bibinfo{author}{\bibfnamefont{B.}~\bibnamefont{Nienhuis}},
  \bibinfo{journal}{Phys. Rev. Lett.} \textbf{\bibinfo{volume}{49}},
  \bibinfo{pages}{1062} (\bibinfo{year}{1982}),
  \urlprefix\url{http://link.aps.org/doi/10.1103/PhysRevLett.49.1062}.

\bibitem[{\citenamefont{Cardy}(2000)}]{Cardy2000}
\bibinfo{author}{\bibfnamefont{J.}~\bibnamefont{Cardy}},
  \bibinfo{journal}{Phys. Rev. Lett.} \textbf{\bibinfo{volume}{84}},
  \bibinfo{pages}{3507} (\bibinfo{year}{2000}),
  \urlprefix\url{http://link.aps.org/doi/10.1103/PhysRevLett.84.3507}.

\bibitem[{\citenamefont{Jacobsen et~al.}(2003)\citenamefont{Jacobsen, Read, and
  Saleur}}]{Jacobsen2003}
\bibinfo{author}{\bibfnamefont{J.~L.} \bibnamefont{Jacobsen}},
  \bibinfo{author}{\bibfnamefont{N.}~\bibnamefont{Read}}, \bibnamefont{and}
  \bibinfo{author}{\bibfnamefont{H.}~\bibnamefont{Saleur}},
  \bibinfo{journal}{Phys. Rev. Lett.} \textbf{\bibinfo{volume}{90}},
  \bibinfo{pages}{090601} (\bibinfo{year}{2003}),
  \urlprefix\url{http://link.aps.org/doi/10.1103/PhysRevLett.90.090601}.

\bibitem[{\citenamefont{FLOHR}(2003)}]{Flohr2003}
\bibinfo{author}{\bibfnamefont{M.~A.~I.} \bibnamefont{FLOHR}},
  \bibinfo{journal}{International Journal of Modern Physics A}
  \textbf{\bibinfo{volume}{18}}, \bibinfo{pages}{4497} (\bibinfo{year}{2003}).

\bibitem[{\citenamefont{{Miller} and {Miyake}}(2015)}]{2015arXiv150802695M}
\bibinfo{author}{\bibfnamefont{J.}~\bibnamefont{{Miller}}} \bibnamefont{and}
  \bibinfo{author}{\bibfnamefont{A.}~\bibnamefont{{Miyake}}},
  \bibinfo{journal}{ArXiv e-prints}  (\bibinfo{year}{2015}),
  \eprint{1508.02695}.

\bibitem[{\citenamefont{Jain}(2007)}]{Jain2007}
\bibinfo{author}{\bibfnamefont{J.}~\bibnamefont{Jain}},
  \emph{\bibinfo{title}{Composite Fermions}} (\bibinfo{publisher}{Cambridge
  University Press}, \bibinfo{year}{2007}), ISBN \bibinfo{isbn}{9781139462648},
  \urlprefix\url{http://books.google.co.il/books?id=0jv9UF6UL20C}.

\bibitem[{\citenamefont{Girvin and MacDonald}(1987)}]{Girvin1987}
\bibinfo{author}{\bibfnamefont{S.~M.} \bibnamefont{Girvin}} \bibnamefont{and}
  \bibinfo{author}{\bibfnamefont{A.~H.} \bibnamefont{MacDonald}},
  \bibinfo{journal}{Phys. Rev. Lett.} \textbf{\bibinfo{volume}{58}},
  \bibinfo{pages}{1252} (\bibinfo{year}{1987}),
  \urlprefix\url{http://link.aps.org/doi/10.1103/PhysRevLett.58.1252}.

\bibitem[{\citenamefont{May}(1967)}]{Plasma1}
\bibinfo{author}{\bibfnamefont{R.}~\bibnamefont{May}},
  \bibinfo{journal}{Physics Letters A} \textbf{\bibinfo{volume}{25}},
  \bibinfo{pages}{282 } (\bibinfo{year}{1967}), ISSN \bibinfo{issn}{0375-9601},
  \urlprefix\url{http://www.sciencedirect.com/science/article/pii/0375960167909061}.

\bibitem[{\citenamefont{Knorr}(1968)}]{Plasma2}
\bibinfo{author}{\bibfnamefont{G.}~\bibnamefont{Knorr}},
  \bibinfo{journal}{Physics Letters A} \textbf{\bibinfo{volume}{28}},
  \bibinfo{pages}{166 } (\bibinfo{year}{1968}), ISSN \bibinfo{issn}{0375-9601},
  \urlprefix\url{http://www.sciencedirect.com/science/article/pii/0375960168904519}.

\bibitem[{\citenamefont{Hansen and Viot}(1983)}]{Plasma3}
\bibinfo{author}{\bibfnamefont{J.}~\bibnamefont{Hansen}} \bibnamefont{and}
  \bibinfo{author}{\bibfnamefont{P.}~\bibnamefont{Viot}},
  \bibinfo{journal}{Physics Letters A} \textbf{\bibinfo{volume}{95}},
  \bibinfo{pages}{155 } (\bibinfo{year}{1983}), ISSN \bibinfo{issn}{0375-9601},
  \urlprefix\url{http://www.sciencedirect.com/science/article/pii/0375960183908228}.

\bibitem[{\citenamefont{Kosterlitz and Thouless}(1973{\natexlab{a}})}]{Plasma4}
\bibinfo{author}{\bibfnamefont{J.~M.} \bibnamefont{Kosterlitz}}
  \bibnamefont{and} \bibinfo{author}{\bibfnamefont{D.~J.}
  \bibnamefont{Thouless}}, \bibinfo{journal}{Journal of Physics C: Solid State
  Physics} \textbf{\bibinfo{volume}{6}}, \bibinfo{pages}{1181}
  (\bibinfo{year}{1973}{\natexlab{a}}),
  \urlprefix\url{http://stacks.iop.org/0022-3719/6/i=7/a=010}.

\bibitem[{\citenamefont{Bonderson et~al.}(2011)\citenamefont{Bonderson,
  Gurarie, and Nayak}}]{Bonderson2011}
\bibinfo{author}{\bibfnamefont{P.}~\bibnamefont{Bonderson}},
  \bibinfo{author}{\bibfnamefont{V.}~\bibnamefont{Gurarie}}, \bibnamefont{and}
  \bibinfo{author}{\bibfnamefont{C.}~\bibnamefont{Nayak}},
  \bibinfo{journal}{Phys. Rev. B} \textbf{\bibinfo{volume}{83}},
  \bibinfo{pages}{075303} (\bibinfo{year}{2011}),
  \urlprefix\url{http://link.aps.org/doi/10.1103/PhysRevB.83.075303}.

\bibitem[{\citenamefont{Cardy}(1986)}]{Cardy1986}
\bibinfo{author}{\bibfnamefont{J.~L.} \bibnamefont{Cardy}},
  \bibinfo{journal}{Nuclear Physics B} \textbf{\bibinfo{volume}{270}},
  \bibinfo{pages}{186 } (\bibinfo{year}{1986}), ISSN \bibinfo{issn}{0550-3213},
  \urlprefix\url{http://www.sciencedirect.com/science/article/pii/0550321386905523}.

\bibitem[{\citenamefont{Freedman et~al.}(2004)\citenamefont{Freedman, Nayak,
  Shtengel, Walker, and Wang}}]{freedman2004class}
\bibinfo{author}{\bibfnamefont{M.}~\bibnamefont{Freedman}},
  \bibinfo{author}{\bibfnamefont{C.}~\bibnamefont{Nayak}},
  \bibinfo{author}{\bibfnamefont{K.}~\bibnamefont{Shtengel}},
  \bibinfo{author}{\bibfnamefont{K.}~\bibnamefont{Walker}}, \bibnamefont{and}
  \bibinfo{author}{\bibfnamefont{Z.}~\bibnamefont{Wang}},
  \bibinfo{journal}{Annals of Physics} \textbf{\bibinfo{volume}{310}},
  \bibinfo{pages}{428} (\bibinfo{year}{2004}).

\bibitem[{\citenamefont{Fendley}(2008)}]{Fendley20083113}
\bibinfo{author}{\bibfnamefont{P.}~\bibnamefont{Fendley}},
  \bibinfo{journal}{Annals of Physics} \textbf{\bibinfo{volume}{323}},
  \bibinfo{pages}{3113 } (\bibinfo{year}{2008}), ISSN
  \bibinfo{issn}{0003-4916},
  \urlprefix\url{http://www.sciencedirect.com/science/article/pii/S0003491608000614}.

\bibitem[{\citenamefont{Gu and Wen}(2014)}]{PhysRevB.90.115141}
\bibinfo{author}{\bibfnamefont{Z.-C.} \bibnamefont{Gu}} \bibnamefont{and}
  \bibinfo{author}{\bibfnamefont{X.-G.} \bibnamefont{Wen}},
  \bibinfo{journal}{Phys. Rev. B} \textbf{\bibinfo{volume}{90}},
  \bibinfo{pages}{115141} (\bibinfo{year}{2014}),
  \urlprefix\url{http://link.aps.org/doi/10.1103/PhysRevB.90.115141}.

\bibitem[{\citenamefont{{Perez-Garcia}
  et~al.}(2007)\citenamefont{{Perez-Garcia}, {Verstraete}, {Cirac}, and
  {Wolf}}}]{Perez-Garcia2007}
\bibinfo{author}{\bibfnamefont{D.}~\bibnamefont{{Perez-Garcia}}},
  \bibinfo{author}{\bibfnamefont{F.}~\bibnamefont{{Verstraete}}},
  \bibinfo{author}{\bibfnamefont{J.~I.} \bibnamefont{{Cirac}}},
  \bibnamefont{and} \bibinfo{author}{\bibfnamefont{M.~M.}
  \bibnamefont{{Wolf}}}, \bibinfo{journal}{ArXiv e-prints}
  (\bibinfo{year}{2007}), \eprint{0707.2260}.

\bibitem[{\citenamefont{{Verstraete} and {Cirac}}(2004)}]{Verstraete2004}
\bibinfo{author}{\bibfnamefont{F.}~\bibnamefont{{Verstraete}}}
  \bibnamefont{and} \bibinfo{author}{\bibfnamefont{J.~I.}
  \bibnamefont{{Cirac}}}, \bibinfo{journal}{eprint arXiv:cond-mat/0407066}
  (\bibinfo{year}{2004}), \eprint{cond-mat/0407066}.

\bibitem[{\citenamefont{Kosterlitz and Thouless}(1973{\natexlab{b}})}]{KT}
\bibinfo{author}{\bibfnamefont{J.~M.} \bibnamefont{Kosterlitz}}
  \bibnamefont{and} \bibinfo{author}{\bibfnamefont{D.~J.}
  \bibnamefont{Thouless}}, \bibinfo{journal}{Journal of Physics C: Solid State
  Physics} \textbf{\bibinfo{volume}{6}}, \bibinfo{pages}{1181}
  (\bibinfo{year}{1973}{\natexlab{b}}),
  \urlprefix\url{http://stacks.iop.org/0022-3719/6/i=7/a=010}.

\bibitem[{\citenamefont{Cheng and Gu}(2014)}]{MengCheng2014}
\bibinfo{author}{\bibfnamefont{M.}~\bibnamefont{Cheng}} \bibnamefont{and}
  \bibinfo{author}{\bibfnamefont{Z.-C.} \bibnamefont{Gu}},
  \bibinfo{journal}{Phys. Rev. Lett.} \textbf{\bibinfo{volume}{112}},
  \bibinfo{pages}{141602} (\bibinfo{year}{2014}),
  \urlprefix\url{http://link.aps.org/doi/10.1103/PhysRevLett.112.141602}.

\bibitem[{\citenamefont{Tsvelik}(2001)}]{Tsvelik2001479}
\bibinfo{author}{\bibfnamefont{A.}~\bibnamefont{Tsvelik}},
  \bibinfo{journal}{Nuclear Physics B} \textbf{\bibinfo{volume}{612}},
  \bibinfo{pages}{479 } (\bibinfo{year}{2001}), ISSN \bibinfo{issn}{0550-3213},
  \urlprefix\url{http://www.sciencedirect.com/science/article/pii/S0550321301003340}.

\bibitem[{\citenamefont{Wang et~al.}(2015{\natexlab{b}})\citenamefont{Wang,
  Santos, and Wen}}]{Juven2}
\bibinfo{author}{\bibfnamefont{J.~C.} \bibnamefont{Wang}},
  \bibinfo{author}{\bibfnamefont{L.~H.} \bibnamefont{Santos}},
  \bibnamefont{and} \bibinfo{author}{\bibfnamefont{X.-G.} \bibnamefont{Wen}},
  \bibinfo{journal}{Phys. Rev. B} \textbf{\bibinfo{volume}{91}},
  \bibinfo{pages}{195134} (\bibinfo{year}{2015}{\natexlab{b}}),
  \urlprefix\url{http://link.aps.org/doi/10.1103/PhysRevB.91.195134}.

\bibitem[{\citenamefont{Ringel and Kraus}(2011)}]{Ringel2011}
\bibinfo{author}{\bibfnamefont{Z.}~\bibnamefont{Ringel}} \bibnamefont{and}
  \bibinfo{author}{\bibfnamefont{Y.~E.} \bibnamefont{Kraus}},
  \bibinfo{journal}{Phys. Rev. B} \textbf{\bibinfo{volume}{83}},
  \bibinfo{pages}{245115} (\bibinfo{year}{2011}),
  \urlprefix\url{http://link.aps.org/doi/10.1103/PhysRevB.83.245115}.

\bibitem[{\citenamefont{Kalmeyer and Laughlin}(1987)}]{PhysRevLett.59.2095}
\bibinfo{author}{\bibfnamefont{V.}~\bibnamefont{Kalmeyer}} \bibnamefont{and}
  \bibinfo{author}{\bibfnamefont{R.~B.} \bibnamefont{Laughlin}},
  \bibinfo{journal}{Phys. Rev. Lett.} \textbf{\bibinfo{volume}{59}},
  \bibinfo{pages}{2095} (\bibinfo{year}{1987}),
  \urlprefix\url{http://link.aps.org/doi/10.1103/PhysRevLett.59.2095}.

\end{thebibliography}

\end{document}